\documentclass[twocolumn]{aastex63} 

\usepackage[normalem]{ulem}
\usepackage[version=4]{mhchem}
\usepackage{longtable}

\newcommand{\rev}[1]{\textcolor{black}{#1}}

\shorttitle{Cold water from ice sublimation in EX~Lup}
\shortauthors{Smith et al.}

\newcommand\konkoly{Konkoly Observatory, HUN-REN Research Centre for Astronomy and Earth Sciences, MTA Centre of Excellence, Konkoly-Thege Mikl\'os \'ut 15-17, 1121 Budapest, Hungary}
\newcommand\elte{Institute of Physics and Astronomy, ELTE E\"otv\"os Lor\'and University, P\'azm\'any P\'eter s\'et\'any 1/A, 1117 Budapest, Hungary}
\newcommand\vienna{Department of Astrophysics, University of Vienna, T\"urkenschanzstr. 17, A-1180 Vienna, Austria}
\newcommand\mpia{Max Planck Institute for Astronomy, K\"onigstuhl 17, 69117 Heidelberg, Germany}

\begin{document}

\title{JWST's sharper view of EX~Lup: cold water from ice sublimation during accretion outbursts}

\correspondingauthor{Andrea Banzatti}
\email{banzatti@txstate.edu}

\author[0009-0008-5780-4217]{Sarah A. Smith}
\affil{Department of Physics, Texas State University, 749 N Comanche Street, San Marcos, TX 78666, USA}

\author[0000-0001-7152-9794]{Carlos E. Romero-Mirza}
\affiliation{Center for Astrophysics $\vert$ Harvard \& Smithsonian, Cambridge, MA 02138, USA}

\author[0000-0003-4335-0900]{Andrea Banzatti}
\affil{Department of Physics, Texas State University, 749 N Comanche Street, San Marcos, TX 78666, USA}

\author[0000-0003-1817-6576]{Christian Rab}
\affiliation{University Observatory, Faculty of Physics, Ludwig-Maximilians-Universität München, Scheinerstr. 1, D-81679 Munich}
\affiliation{Max-Planck-Institut für extraterrestrische Physik, Giessenbachstrasse 1, D-85748 Garching, Germany}

\author[0000-0001-6015-646X]{P\'eter \'Abrah\'am}
\affiliation{\konkoly{}}
\affiliation{\elte{}}
\affiliation{\vienna{}}

\author[0000-0001-7157-6275]{\'Agnes K\'osp\'al}
\affiliation{\konkoly{}}
\affiliation{\elte{}}
\affiliation{\mpia{}}

\author[0000-0001-8194-4238]{Rik Claes}
\author[0000-0003-3562-262X]{Carlo F. Manara}
\affil{European Southern Observatory, Karl-Schwarzschild-Strasse 2,
85748 Garching bei München, Germany}

\author[0000-0001-8798-1347]{Karin I. \"Oberg}
\affiliation{Center for Astrophysics, Harvard \& Smithsonian, 60 Garden St., Cambridge, MA 02138, USA}

\author[0000-0003-4757-2500]{Jeroen Bouwman}
\affiliation{\mpia{}}

\author[0000-0002-4283-2185]{Fernando Cruz-S\'aenz de Miera}
\affiliation{\konkoly{}}
\affiliation{Institut de Recherche en Astrophysique et Plan\'etologie, Universit\'e de Toulouse, UT3-PS, OMP, CNRS, 9 av. du Colonel-Roche, 31028 Toulouse Cedex 4, France}

\author[0000-0003-1665-5709]{Joel D. Green}
\affiliation{Space Telescope Science Institute, 3700 San Martin Drive, Baltimore, MD 02138, USA}

\begin{abstract}
The unstable accretion phases during pre-main-sequence evolution of T~Tauri stars produce variable irradiation and heating of planet-forming regions. A strong accretion outburst was observed with Spitzer-IRS in 2008 in EX~Lup, the prototype of EXor variables, and found to increase the mid-infrared water and OH emission and decrease organic emission, suggesting large chemical changes. We present here two JWST-MIRI epochs of quiescent EX~Lup in 2022 and 2023 obtained over a decade after the 2008 outburst and several months after a moderate burst in 2022. With JWST's sharper spectral view, we can now analyze water emission as a function of temperature in the two MIRI epochs and, approximately, also in the previous Spitzer epochs. This new analysis shows a strong cold water vapor ``burst" in low-energy lines during the 2008 outburst, which we consider clear evidence for enhanced ice sublimation due to a recession of the snowline, as found in protostellar envelopes. JWST shows that EX~Lup still has an unusually strong emission from cold water in comparison to other T~Tauri disks, \rev{suggesting $> 10$-yr-long freeze-out timescales in the inner disk surface}. EX~Lup demonstrates that outbursts can significantly change the observed organic-to-water ratios and increase the cold water reservoir, providing chemical signatures to study the recent accretion history of disks. This study provides an unprecedented demonstration of the chemical evolution triggered by accretion outbursts in the Class II phase and of the high potential of time-domain experiments to reveal processes that may have fundamental implications on planet-forming bodies near the snowline.
\end{abstract}


\section{Introduction} \label{sec: intro}
\begin{figure*}
    \centering
    \includegraphics[width=1\textwidth]{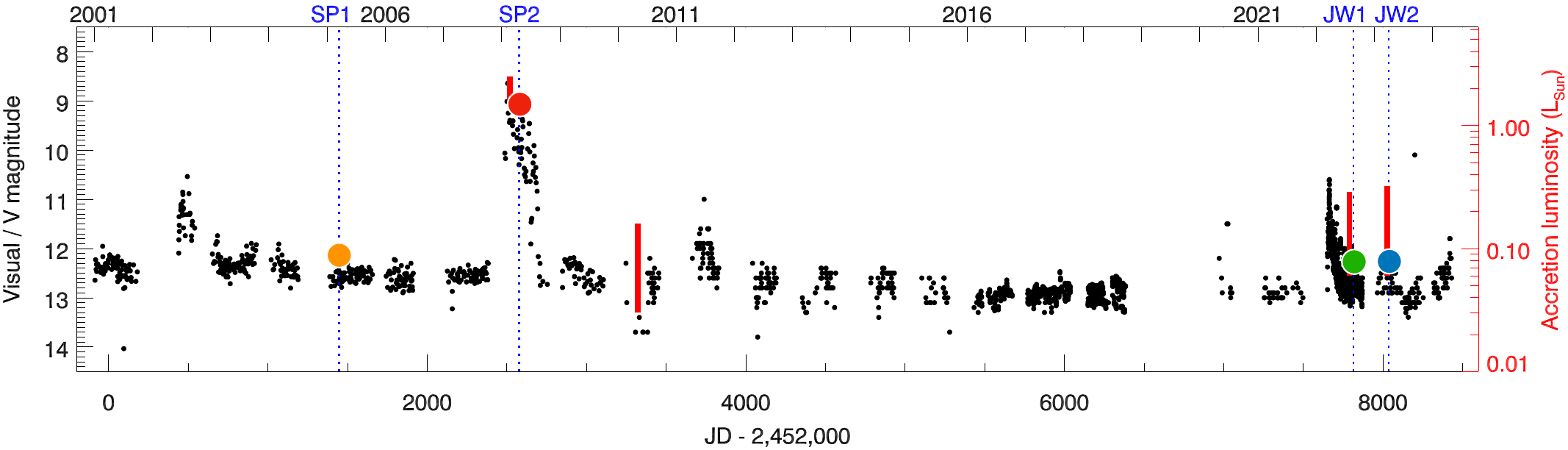}
    \caption{Light curve of EX Lupi during the period containing the four epochs of analysis (SP1 and SP2 from Spitzer-IRS, JW1 and JW2 from JWST-MIRI). V-band data (y axis to the left) are taken from: AAVSO (www.aavso.org), the ASAS-3 catalog \citep{Pojmanski02}, and ASAS-SN \citep{Shappee14,Kochanek17}. Accretion luminosity estimates (y axis to the right) are reported with red bars where available from previous work, and large dots show the values adopted in this work (see Table \ref{tab: epochs} and Section \ref{sec: accr} for details), color-coded to distinguish them in all figures in this paper.}
    \label{fig: light-curve}
\end{figure*}

The chemical composition of gas and dust in protoplanetary disks, as well as their evolution, are proposed to shape planet formation in multiple ways. Molecular snowlines and icy “pebbles” are considered fundamental in the formation of planetary cores and to deliver volatile ices to inner disks where super-Earths and small rocky planets are forming \citep[e.g.][]{stevenson88,cieslacuzzi06,lambrechts12,bitsch19_rocky,izidoro21,drazkowska17}. A direct link between disks and planets is also expected from planetary gas accretion, where the disk gas chemistry near the orbital distance of a planet will determine the elemental ratios of its primary atmosphere \citep[e.g.][]{madhusudhan16,bitsch19_giant,cridland19,krijt23}. 

Warm molecular gas is observed at infrared wavelengths, and previous spectroscopy surveys found that \ce{H2O}, CO, OH, and some small organic molecules (\ce{HCN}, \ce{C2H2} and \ce{CO2}) are common within $\sim 1$~au in protoplanetary disks of T~Tauri stars \citep{cn08,cn11,salyk08,salyk11_spitz,pascucci09,pont10}. These studies found that inner disk molecular abundances show evidence for significant reprocessing from prestellar phases \citep{pont14}, supporting theoretical ideas that disk gas-phase chemistry is fundamental for the evolution of planet-forming regions \citep[e.g.][]{glassgold09}. Solar system material indeed shows evidence for partial or near complete chemical reprocessing within a few au of the Sun, as traced by meteoritic material \citep{grossman72}. 

While a hot inner disk in itself may provide conditions for chemical reset \citep{pont14,bosman17}, young stars are known to have episodic, strongly variable accretion rates that inject high-energy radiation and heat into the inner disk through events of episodic outbursts, where the accretion luminosity (mostly UV radiation) increases between factors of a few up to factors of $\sim 100$ throughout the pre-main-sequence evolution from protostar to star \citep[see reviews by][]{hartmann96,audard14,fischer23}. These accretion bursts are expected to impact the inner disk chemistry significantly, since UV and X-ray radiation play a key role in the heating, excitation, and photo-chemistry of molecules \citep[e.g.][]{vorobyov13,walsh15,woitke18,du14,molyarova18,najita17,cleeves17}.

The ubiquity and frequency of accretion bursts during star formation at 0.1--1~Myr suggests that burst-driven chemical evolution may constitute a fundamental ingredient in inner disk chemistry, and in turn planet formation. Previous work supported this scenario by studying strong accretion outbursts in embedded young stellar objects that may retain long-term chemical effects in the envelope, or in outer disk regions at 10--100~au \citep[e.g.][]{visser15,cleeves17,rab17,molyarova18,waggoner19}. Observations of Class O/I protostars and FUors objects have indeed shown evidence in multiple objects for outburst-induced sublimation of ices leaving gas-phase molecules beyond their quiescent snowline locations \citep[e.g.][and references therein]{hsieh19,jorgensen20,fischer23,calahan24}. However, the erratic occurrence and lower intensity of these bursts in the Class II phase has made it difficult to design effective monitoring studies of molecules in the inner disk region ($< 5$~au) around classical T~Tauri stars, often limited to a few epochs of spectra \citep{goto11,banz12,banz14,banz15}.

\begin{figure*}
\centering
\includegraphics[width=0.85\textwidth]{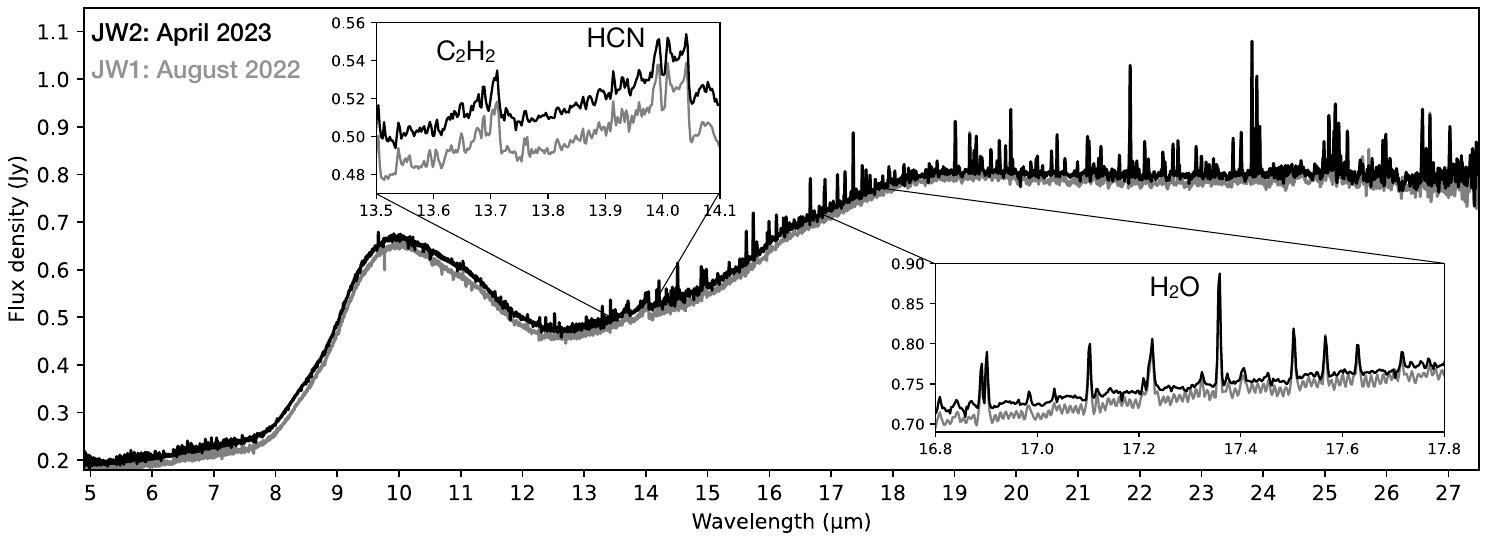} 
\caption{MIRI epochs of EX~Lup analyzed in this work, zooming-in on two regions that illustrate organic and water emission (insets). The 2022 spectrum has been published before in \cite{kospal23} and here we provide the new reduction, where the improved spectral response function removes spurious broad features previously attributed to dust emission \citep[compare to Figures 1 and 2 in][]{kospal23}. The 2022 epoch also had higher residual fringes (see inset) due to a large pointing offset. \rev{The new reduction of both spectra is available on SpExoDisks.com \citep{SpExoDisks}.}}
\label{fig: MIRI_original}
\end{figure*}

So far, large changes in inner disk chemistry triggered by accretion outbursts have been observed in at least one Class II system: EX~Lup, an M0 star of $\sim 1$~Myr at a distance of 155~pc \citep{gaia_mission,gaiaDR3} and the prototype for EXor outbursts, which are thought to happen later during pre-stellar evolution than the stronger FUor outbursts \citep{herbig08,fischer23}.
EX~Lup had a strong outburst in 2008, when the accretion luminosity increased by a factor $\approx$~40--100 \citep[][and Figure \ref{fig: light-curve}]{aspin10,wang23,cruzsaenzdemiera23}. Comparison of mid-infrared (10--37~$\mu$m) Spitzer-InfraRed Spectrograph \citep[IRS,][]{irs} spectra taken before and during the outburst showed strong changes in the continuum level and in molecular emission that were interpreted in the context of inner disk heating and UV photo-dissociation of molecules: the dust continuum increased by a factor of 4--5, organic emission disappeared, highly-excited OH emission increased as a water dissociation product, and water emission increased at all wavelengths from 3 to 35 $\mu$m \citep[][]{banz12,banz15}. Unfortunately, the low resolving power of Spitzer-IRS spectra ($\sim 450$~km/s, which blends together transitions from different energy levels and molecules) limited and still limits the interpretation of all these changes. No follow-up studies of the mid-infrared gas emission after outburst were possible due to Spitzer entering the warm phase in 2009 \citep{juhasz12}. Only with the launch of JWST has the community regained access to the rich molecular disk spectra at mid-infrared wavelengths. Re-observation of EX Lup in Cycle 1, 14 years after the 2008 outburst, showed that the inner disk organic molecules are detected again \citep{kospal23}.

The case of EX Lup suggests a fundamental role for variable UV irradiation and disk heating in inner disk chemistry, which might contribute to the large scatter in molecular line luminosities and organic-to-water ratios measured in samples of T~Tauri disks of similar age \citep[][Arulanantham et al. submitted]{cn11,salyk11_spitz,banz20}. These scatters still challenge disk chemical models due to multiple factors and their degeneracy \citep[dust-to-gas ratios, elemental abundances, dust properties, disk irradiation, e.g.][]{najita11,woitke18,anderson21}. In this work, we analyze molecular emission in two epochs on MIRI spectra of EX Lup obtained in 2022 and 2023 in comparison to the previous IRS epochs published in \citet{banz12} and the samples of T Tauri stars observed with Spitzer-IRS from \citet{banz20} and with MIRI-MRS from \citet{banzatti24}, to investigate the post-outburst chemical evolution observed in EX~Lup and gain a better understanding of the variability observed in the water spectrum in particular.

\section{Observations \& accretion epochs} \label{sec: data}
EX Lup was observed at 4.9--28\,$\mu$m with the Medium Resolution Spectrometer \citep[MRS,][]{jwst-mrs} mode on JWST-MIRI \citep[][]{miri,miri2} as part of programs GO-2209 and GO-4427 (a follow-up DDT program) in Cycle 1 (PI: P. \'Abrah\'am), with the primary goal to monitor the evolution in dust and gas emission after the 2008 outburst and a recent moderate burst that happened in March 2022 \citep[][and Figure \ref{fig: light-curve}]{cruzsaenzdemiera23}. The two epochs of MIRI data are shown in Figure \ref{fig: MIRI_original}: the first was obtained on August 23, 2022 (labeled JW1 in this work) and the second on April 3, 2023 (labeled JW2). Except for a minor difference in continuum flux, the observed molecular emission is very similar in the two MIRI epochs with prominent emission from four carbon-bearing molecules that are common in disks (CO, \ce{HCN}, \ce{C2H2}, and \ce{CO2}), OH, and with a water spectrum that shows prominent low-energy lines (see more in Section \ref{sec: analysis}). 

A pointing offset of $\sim 1''$ in the first MIRI epoch (JW1), which was obtained without target acquisition, resulted in a lower-quality spectrum with residual fringes (see e.g. around 17~$\mu$m in Figure \ref{fig: MIRI_original}) and partial loss of data. The second epoch (JW2) was therefore taken with target acquisition, resulting in a higher S/N spectrum with much improved fringe correction. The first epoch has been published in \citet{kospal23}; in this work, we present an improved data reduction (Appendix \ref{app: reduction}) that removes artifacts and residuals present from a non-optimal characterization of the spectral response function, which caused ``bumps" that were interpreted as dust emission in \citet{kospal23}. 

\begin{figure*}
\centering
\includegraphics[width=1\textwidth]{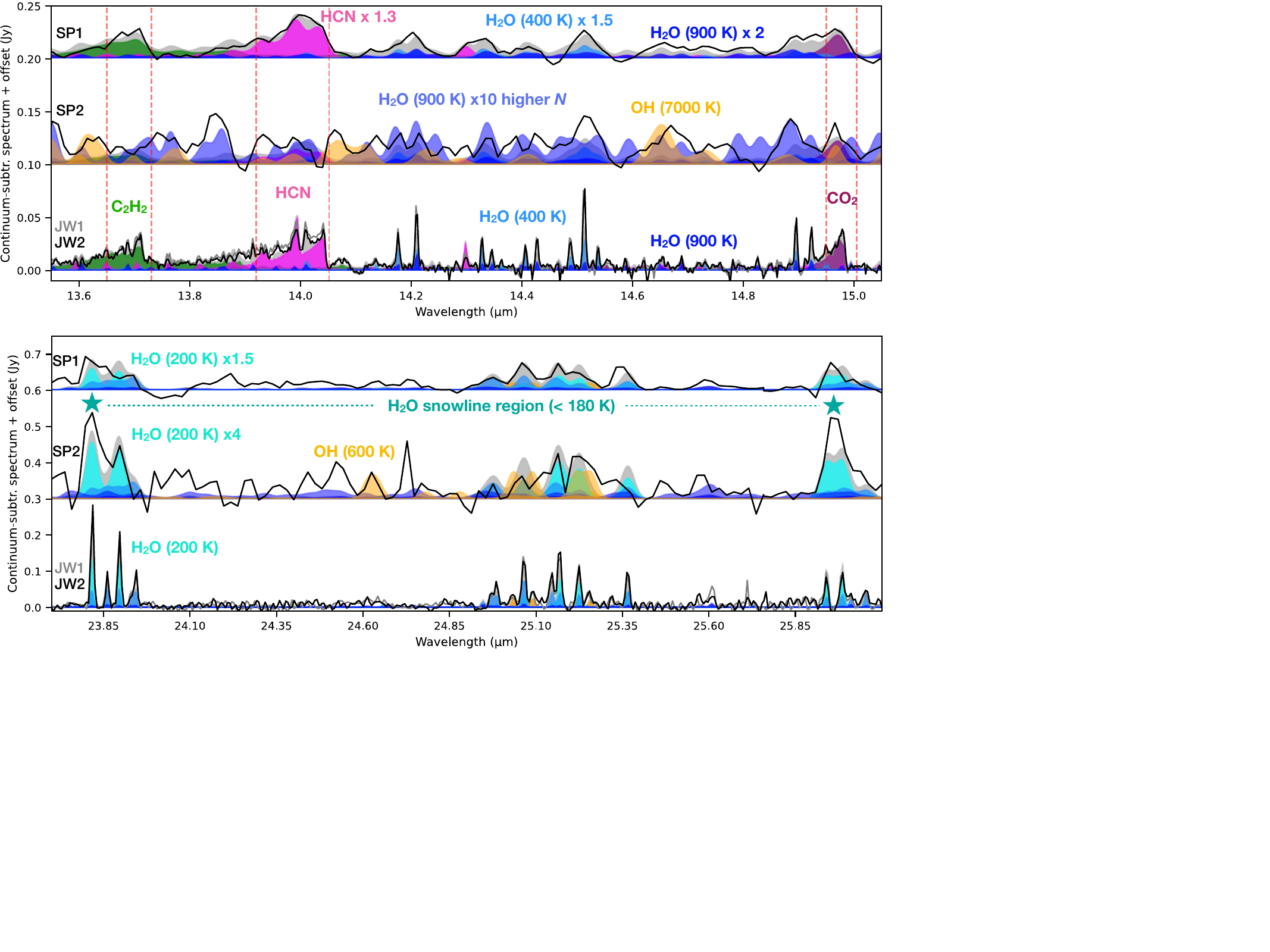} 
\caption{Mid-infrared spectral line variability observed with Spitzer-IRS \citep{banz12} and JWST-MIRI (this work) in the T Tauri disk of EX Lup over $\sim 20$ years (from 2005 with the SP1 epoch, to 2023 with the JW2 epoch). The two MIRI epochs are plotted in grey (JW1) and black (JW2) without offset, to demonstrate their similarity. The individual molecular models obtained for the JW2 epoch (Table \ref{tab: slab results}) are shown in different colors, their sum in grey. The same models are shown as downgraded to the resolving power of Spitzer-IRS for comparison to the SP1 and SP2 data, \rev{with multiplication factors as labeled for each molecule;} additional models for water and OH are shown on top of the SP2 epoch to explain the more complex structure observed (see text for details). The vertical dashed lines show the range where the line flux of organic molecules is measured for Figure \ref{fig: IRS_correl}. \rev{The lowest-energy lines covered by MIRI ($E_u$ = 878--1448~K), which are more sensitive to the snowline region at T $< 180~K$, are marked with a star (see also Figure \ref{fig: 23umlines_ratios}).}
}
\label{fig: all_spectra}
\end{figure*}

The two MIRI epochs are compared in this work to two earlier spectra from Spitzer IRS that were already published in \citet{banz12}, from 2005 and 2008 (Figure \ref{fig: light-curve}, and data reduction description in the original paper). Before the analysis of the molecular spectra in each epoch from Spitzer and JWST, we removed the dust continuum using the procedure described in \citet{pontoppidan24} and \citet{banzatti24}; in the IRS spectra, this procedure mostly captures a pseudo continuum made of weak emission lines, as discussed before \citep{salyk11_spitz}. The four epochs are listed in Table \ref{tab: epochs} and the main spectral regions of interest are shown in Figure \ref{fig: all_spectra}.

\subsection{Accretion estimates and variability} \label{sec: accr}
The accretion rates and luminosities in the different epochs of EX~Lup spectra are important for the interpretation of the observed molecular spectra and their variability. Unfortunately, simultaneous accretion estimates are not available from other works or datasets for any of the epochs included here and we have to adopt approximate values as follows. The values available from previous works are shown as vertical red bars (showing their uncertainty) in Figure \ref{fig: light-curve}, while the values we adopt are shown as large colored dots in each epoch and listed in Table \ref{tab: epochs}.
Some estimates of accretion luminosities in different epochs of EX~Lup are available from fits to the Balmer Jump by \cite{cruzsaenzdemiera23} and \citet{wang23} from spectra taken with the X-Shooter spectrograph on the Very Large Telescope \citep{xshooter}. These two works used two different values for the extinction $A_{\rm{V}}$ = 0.1 and 1.1 mag, which give a different accretion luminosity by a factor $\sim 5$ even when using the same exact spectrum \citep[see discussion in][]{cruzsaenzdemiera23,wang23}. This range gives an idea of the spread of values due to uncertainties in $A_{\rm{V}}$, which can generally contribute to the spread of accretion luminosities estimated in other disks too (see Section \ref{sec: analysis}).

For the Spitzer epoch in 2005, we use the accretion value from a Balmer Jump fit to the X-Shooter spectrum from May 4th, 2010. We assume the accretion rate in 2010 approximately represents the quiescent epoch in 2005; the $V$-band photometry was indeed similar, within 0.5 mag (see Figure \ref{fig: light-curve}). For the 2008 epoch, we take the strong HI 7-6 line observed in the Spitzer spectrum (log(HI/$L_{\odot}$) = -4.6) and convert that into an accretion luminosity using the relations from \citet{Tofflemire25}.
For the JWST epochs, there is an accretion estimate available from a fit to the Balmer Jump in X-Shooter data obtained a month before epoch JW1 on July 29, 2022 from \citet{cruzsaenzdemiera23}, and we fit the Balmer Jump in X-Shooter data obtained a week before JW2 on March 24, 2023 (this spectrum will be presented in Kospal et al. in prep.) following methods described in \cite{manara13,Claes24}. 

\begin{deluxetable}{l l c c c}
\tabletypesize{\footnotesize}
\tablewidth{0pt}
\tablecaption{\label{tab: epochs} EX Lup epochs included in this work.}
\tablehead{ Epoch & Date & Instr. & log$L_{\rm{acc}}/L_{\odot}$ & Ref.}
\tablecolumns{6}
\startdata
SP1 & 2005 Mar 18 & IRS & -1.02(0.23)  & a, b \\
SP2 & 2008 Apr 21 & IRS & 0.2(0.3) &  c \\
JW1 & 2022 Aug 23 & MIRI & -1.11(0.11) & c \\
JW2 & 2023 Apr 3 & MIRI & -1.11(0.11)  & c \\
\enddata
\tablecomments{Accretion values adopted for each epoch (see Figure \ref{fig: light-curve} and Section \ref{sec: accr}): a- \cite{wang23}; b- \cite{cruzsaenzdemiera23}; c- this work. Accretion for the quiescent phase in 2005 is assumed from 2010.}
\end{deluxetable} 

Despite the time lag between the accretion measurements and the MIRI data, the fits to X-shooter data suggest that the accretion was approximately the same in the two MIRI epochs, in the range of 0.06--0.3 $L_{\odot}$ (using the range of possible extinction $A_{\rm{V}}$ = 0.1 and 1.1 mag, as reported above). From the photometric data in Figure \ref{fig: light-curve}, the visible photometry of EX Lup in August 2022 and April 2023 indeed spans a similar range within 0.5 mag, modulated by the known stellar spots on a 7.4 day period \citep{sicagul23,pouilly24}. Additionally, we use the relations provided by \citet{Tofflemire25} to provide a simultaneous accretion estimate from HI lines observed in the MIRI spectra. The non-detection of HI in JW1 and JW2 (with upper limits of $\lesssim 10^{-6} L_{\odot}$ for both the 10-7 and 7-6 lines) implies $L_{\rm{acc}} \lesssim 0.1$~$L_{\odot}$ for both MIRI epochs, supporting the lower end of the range estimated from X-Shooter in 2022 and 2023. Combining the constraints obtained from X-shooter and MIRI, we adopt an $L_{\rm{acc}}$ range 0.06--0.1~$L_{\odot}$ in this work for both JWST epochs (Table \ref{tab: epochs}). The conclusion that the accretion luminosity did not change significantly between the two MIRI epochs is consistent with the minor variability in their dust and gas emission (Figure \ref{fig: MIRI_original} and Section \ref{sec: analysis}).

\begin{deluxetable}{l c c c c}
\tabletypesize{\small}
\tablewidth{0pt}
\tablecaption{\label{tab: slab results} Slab model results.}
\tablehead{Species & $T$ & $N$ & $A_{\rm{slab}}$ & $M$ \\
 & (K) & (cm$^{-2}$) & (au$^2$) & ($M_{\oplus}$) }
\tablecolumns{5}
\startdata
\ce{H2O} hot & 900$^a$ & $1.9^{+0.4}_{-0.3} \times 10^{18}$ & $0.035^{+0.003}_{-0.003}$ & $7.5 \times 10^{-8}$ \\
\ce{H2O} warm & 400$^a$ & $2.8^{+0.4}_{-0.4} \times 10^{18}$ & $1.3^{+0.1}_{-0.1}$ & $4.1 \times 10^{-6}$ \\
\ce{H2O} cold & 200$^a$ & $2.6^{+0.8}_{-0.7} \times 10^{17}$ & $43^{+6.5}_{-10.1}$ & $1.2 \times 10^{-5}$ \\
\ce{CO2} & $493^{+41}_{-36}$ & $1.0^{+0.6}_{-0.6} \times 10^{17}$ & $0.09^{+0.08}_{-0.02}$ & $2.7 \times 10^{-8}$ \\
\ce{C2H2} & $1294^{+178}_{-106}$ & $4.5^{+2.3}_{-2.3} \times 10^{17}$ & $0.005^{+0.004}_{-0.001}$ & $3.8 \times 10^{-9}$ \\
\ce{HCN} & 692$^{+16}_{-16}$ & ($1.5 \times 10^{14}$)$^b$ & (52)$^b$ & $1.4 \times 10^{-8}$ \\
\ce{OH} & 1309$^{+80}_{-73}$ & ($2.2 \times 10^{13}$)$^b$ & (55)$^b$ & $1.3 \times 10^{-9}$ \\
\enddata
\tablecomments{The best-fit parameter values and their confidence limits are the median and 0.16--0.84 posterior distribution percentiles, following \citet{munozromero24b}. $^a$: temperature fixed. $^b$: solution highly degenerate between $N$ (the column density) and $A_{\rm{slab}}$ (the slab emitting area). The last column reports the observed mass, from the product of area, column density, and molecular mass.}
\end{deluxetable}

\section{Analysis \& Results} \label{sec: analysis}

\subsection{A slab model fit to the MIRI spectrum and its comparison to previous IRS epochs of EX~Lup} \label{sec: slab_model}
To estimate the properties of the emitting gas, we fit the spectrum observed with MIRI with a slab model in local thermodynamic equilibrium (LTE). Given its similarity to JW1 and the lower-quality of that spectrum (Section \ref{sec: data}), we only fit the JW2 epoch. We use the Python package iris \citep{iris} as implemented in \citet{munozromero24,munozromero24b}. In brief, the molecular spectra are defined by an excitation temperature $T$ (in K), column density $N$ (in cm$^{-2}$), a slab emitting area $A_{\rm{slab}} = \pi R_{\rm{slab}}^{2}$ (in au$^{2}$), and a line width due to turbulence (assumed to have Gaussian shape with standard deviation of 1~km/s) plus thermal broadening (which is about 0.4--1 km/s for the temperatures found in EX Lup). The synthetic models are then convolved with the MIRI resolving power in each sub-band as provided in \citet{pontoppidan24} and \citet{banzatti24}.
The code accounts for line blending and line-center saturation by computing opacity-weighted line intensities; this has been found to be important to correctly estimate the column density of organics \citep{tabone23} and water \citep{banzatti24} from MIRI spectra. We simultaneously fit the 12--27~$\mu$m region in the continuum-subtracted spectrum including the following molecules: \ce{H2O}, \ce{OH}, \ce{HCN}, \ce{C2H2}, and \ce{CO2}. Since a temperature gradient is needed to reproduce \ce{H2O} emission across MIRI wavelengths \citep{banz23,banz23b,gasman23,temmink24b}, we include three components of \ce{H2O} with fixed temperatures at 900~K, 400~K, and 200~K as an approximation \citep{munozromero24,banzatti24}.

The best-fit results are reported in Table \ref{tab: slab results} and shown in Figure \ref{fig: all_spectra}. For water, we find similar results to previous work that used multiple temperature components, although it is not possible at this time to compare the estimated masses at different temperatures due to the different approaches taken in different works \citep[number of components and their temperature, see][]{munozromero24b,temmink24b}. Direct comparison of the spectra shows that EX~Lup has an uncommon cold water reservoir, as shown in Appendix \ref{app: MIRI_compar}.
As for the other molecules, we find similar temperatures to those found in other MIRI disk spectra analyzed with single-temperature slab models in LTE, suggesting higher temperatures for OH, HCN, and \ce{C2H2} in comparison to water and \ce{CO2}, and also the typical degeneracies between column density and emitting area \citep[e.g.][Arulanantham et al. submitted]{grant23,gasman23,schwarz24,salyk25}.

We do not attempt to fit the Spitzer spectra due to the large degeneracies that are known to plague slab model results at the low resolution of IRS \citep{salyk11_spitz,banz12,munozromero24}. Instead, in Figure \ref{fig: all_spectra}, we compare the best-fit JW2 model to the Spitzer epochs of data, by convolving the same model down to the IRS resolution. In the quiescent SP1 epoch, the emission is overall well reproduced by the JW2 model but shows stronger water and HCN emission by factors of $\sim$~1.5--2 and 1.3, respectively. These changes are consistent with small variations in the model parameters: emitting area, column density, and/or temperature.  

\begin{figure*}
\centering
\includegraphics[width=1\textwidth]{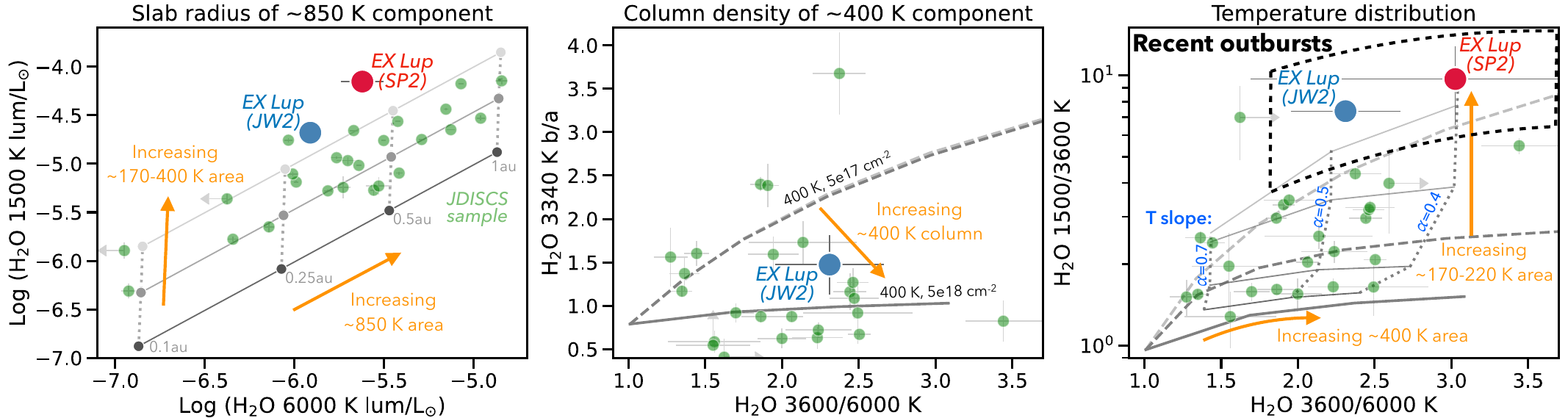} 
\caption{Water diagnostic diagrams from \citet{banzatti24}, with the JDISCS sample from that work shown for reference in light green. The position of the JW2 spectrum of EX~Lup (large blue datapoint) in these diagrams shows: 1) a moderate slab radius ($\approx 0.3$~au) for the hot water component (left plot), 2) optically thick emission ($N \approx 10^{18}$~cm$^{-2}$) in the warm water component (middle plot), and 3) a temperature gradient with negative slope $\approx 0.45$--0.55 and with strong cold water enrichment (right plot, see also Figure \ref{fig: 23umlines_ratios}). These estimates are confirmed by the slab fit results both with discrete components and with the radial gradient (Sections \ref{sec: slab_model} and \ref{sec: cooler}). \rev{The line ratios from the outburst SP2 spectrum are more uncertain due to line blending (see Appendix \ref{app: deblending}), but indicate a larger and colder water reservoir (see also Figure \ref{fig: 23umlines_ratios}).}}
\label{fig: diagnostic_diagram}
\end{figure*}

\begin{figure}
\centering
\includegraphics[width=0.4\textwidth]{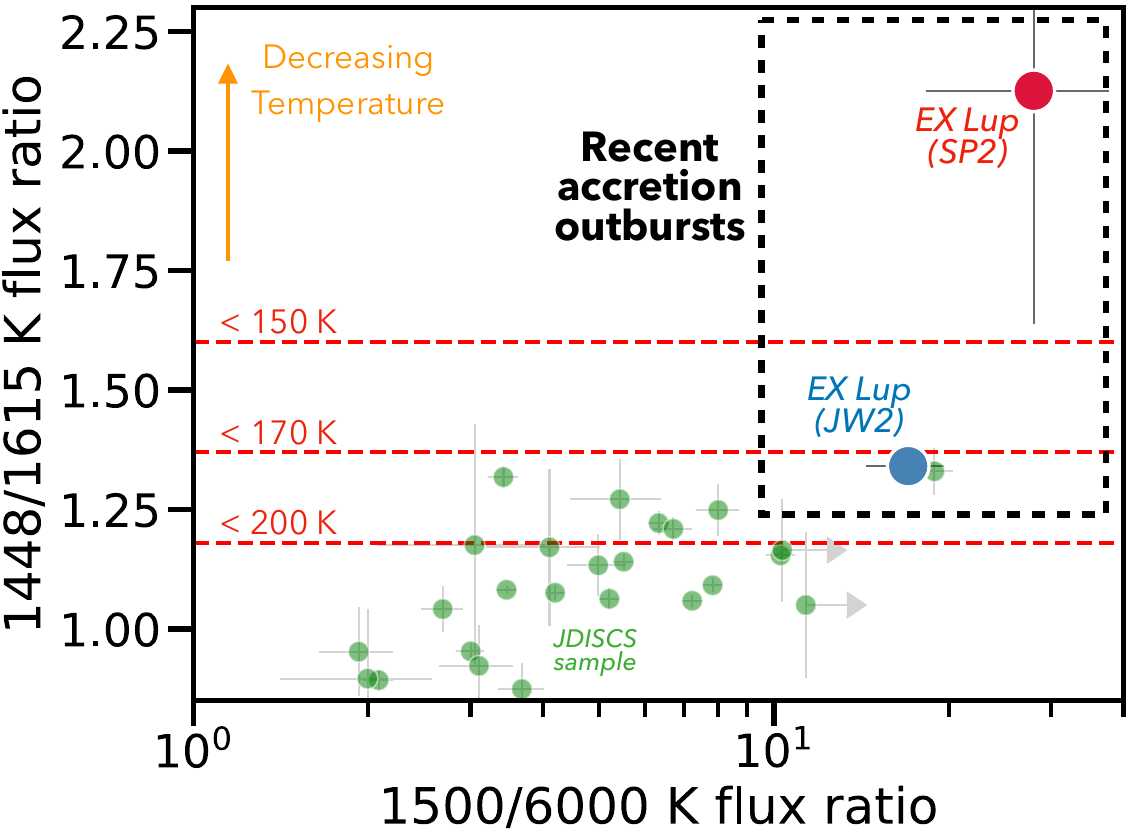} 
\caption{Coldest water detected in EX~Lup, in comparison to the JDISCS sample from \citet{banzatti24}. The 23~$\mu$m line flux asymmetry diagnostic (the 1448/1615~K flux ratio) shows one of the largest values ever measured before, consistent with ice sublimation in the inner disk \citep[\rev{the temperatures associated to specific 1448/1615~K values assume a column density of $N = 10^{17}$~cm$^{-2}$, see}][]{banzatti24}. The other object near EX~Lup (JW2) is the young disk of IRAS~04385-2550.}
\label{fig: 23umlines_ratios}
\end{figure}

In the outburst SP2 epoch, instead, the observed emission structures are very different from quiescence and while water and OH are still strong, the characteristic emission features from \ce{C2H2} and \ce{HCN} are not observed any more, as already reported in \citet{banz12}. While in principle their emission could still be present at lower levels, the structure observed at 13--15~$\mu$m suggest that emission from other species dominates. Following previous analyses of water and OH in Spitzer-IRS and VLT-CRIRES spectra \citep{banz12,banz15}, after including additional hot emission from OH and \ce{H2O} models we do not find any convincing evidence for any detection of \ce{C2H2} and \ce{HCN} emission. Figure \ref{fig: all_spectra} shows that most of the additional structure observed in SP2 (outburst) align with high-energy transitions that are excited in a high-temperature water model ($T \approx 900$~K, $N \approx 10^{19}$~cm$^{-2}$) and in a very hot OH component ($T = 7000$~K, $N = 10^{17}$~cm$^{-2}$) in addition to a colder OH component ($T = 600$~K, $N = 10^{18}$~cm$^{-2}$). The presence of these two different OH components was found in \citet{banz12}, but it is most likely just an approximation of the complex, non-thermal OH excitation following UV photo-dissociation of water \citep{carr14,tabone24}. We remark that the enhanced OH emission in 2008 was found to populate cross-ladder transitions that overlap with \ce{CO2}, making its detection, too, only tentative \citep{banz12}. 

As for the long wavelengths ($> 20 \mu$m) covering colder \ce{H2O} and OH emission from lower-energy lines (Figure \ref{fig: all_spectra}), again the SP1 epoch is well reproduced by the JW2 model (with a factor of $< 2$ change in line flux), while the SP2 outburst epoch clearly shows stronger lines that can be approximately reproduced with a larger emitting area by factors of 3--5. The largest increase in emission is observed in the lowest-energy lines near 23.85~$\mu$m and 26~$\mu$m, indicating increased emission from the region near the ice sublimation front \citep[$< 180$~K, see][]{banz23b,banzatti24}. \rev{We remark that the fixed-temperature 200~K model in Figure \ref{fig: all_spectra} cannot reproduce the stronger line at 23.82~$\mu$m ($E_u = 1448$~K), which indicates temperatures at or below 170~K (see next section).}

\begin{figure*}
\centering
\includegraphics[width=1\textwidth]{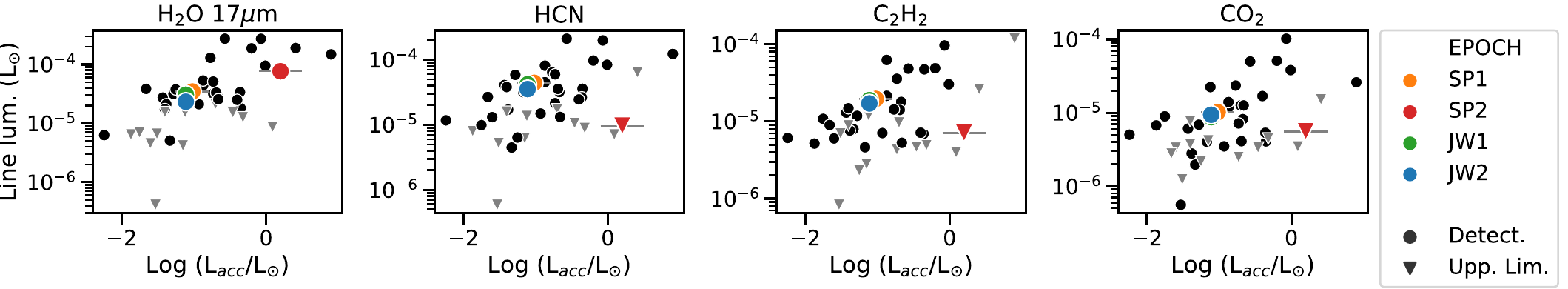} 
\caption{Changes measured in EX Lup in the four epochs included in this work (colored datapoints, using line fluxes from Table \ref{tab: line fluxes}) as compared to molecular luminosities measured in T Tauri disks of similar age from \citet{banz20}.
}
\label{fig: IRS_correl}
\end{figure*}

\subsection{Enhanced cold water in the disk surface of EX Lup} \label{sec: cooler}
Section \ref{sec: slab_model} and Figure \ref{fig: all_spectra} showed strong emission from lower-energy water lines in the spectrum of EX~Lup, both during outburst (with Spitzer) and afterwards (the MIRI epochs). Following \citet{banzatti24}, we now place EX~Lup in the broader context of other T~Tauri disks observed as part of the JDISC Survey \citep[][Arulanantham et al. submitted]{pontoppidan24}. Figures \ref{fig: diagnostic_diagram} and \ref{fig: 23umlines_ratios} show specific water transitions (indicated by their upper level energy in K) and their ratios introduced in \citet{banzatti24} as diagnostics of the water radial distribution in inner disks (see Appendix \ref{app: MIRI_compar}). Transitions from 1500~K levels (observed in MIRI in two strong lines near 23.85~$\mu$m) are sensitive to the colder water reservoir down to snowline temperatures (Figure \ref{fig: 23umlines_ratios}), while higher-energy transitions at 3600~K and 6000~K are sensitive to water-emitting regions at higher temperatures up to $\sim$~1000~K. We measure these lines in the MIRI spectra as well as in the SP2 IRS spectrum, as demonstrated in Appendix \ref{app: deblending}. The interpretation of line ratios in the context of discrete slab models and a power-law temperature gradient is explained in detail in \citet{banzatti24} and summarized in Figure \ref{fig: diagnostic_diagram}.

The position of EX~Lup in the diagrams in Figure \ref{fig: diagnostic_diagram} as compared to the larger disk sample demonstrate three important points: 1) the hot water component in EX~Lup has a moderate emitting area that increases during outburst and more extended warm and cold components than the rest of the sample (left diagram); 2) the warm water component shows optically thick emission ($N \approx 10^{18}$~cm$^{-2}$), similar to other disks (middle diagram); and 3) both the warm and cold components are enhanced in EX~Lup, with ratios that can be described with a temperature gradient with slope $\approx -0.5$ to -0.4 in the two epochs (less steep in outburst). This is further supported by the strong line flux asymmetry measured in the 23.85~$\mu$m lines where the 1448~K line is stronger than the 1615~K line (Figures \ref{fig: all_spectra} and \ref{fig: 23umlines_ratios}), with a ratio that is among the highest ever measured so far and consistent with ice sublimation at $< 180$~K \citep[][]{lodders03}. Overall, EX~Lup shows an uncommonly strong cold water reservoir in comparison to other disks previously observed with MIRI (see also Appendix \ref{app: MIRI_compar}).

As discussed in previous work, discrete temperature components are only an approximation of a temperature gradient in the inner disk \citep{munozromero24b,temmink24b,banzatti24}. As a second step, we therefore apply the power-law temperature profile fits developed in \citet{munozromero24b} to the MIRI JW2 spectrum of EX~Lup, adopting $T_{ex} = T_0 ( r / 0.5 \text{au})^{-a}$ and $N = N_0 ( r / 0.5 \text{au})^{-b}$. As in the other disks analyzed in \citet{munozromero24b} we find an improved fit over the discrete-temperature approximation with these best-fit parameters: $T_0 = 390^{+3}_{-4} K$, $N_0 = 1.6^{+2.4}_{-0.2} \times 10^{18}$~cm$^{-2}$, $a = 0.42^{+0.1}_{-0.1}$, $b = 1.67^{+0.13}_{-0.10}$. The best fit agrees very well with the approximate estimates described above based on the diagnostic diagrams in Figure \ref{fig: diagnostic_diagram}, supporting their interpretation in the framework of radial gradients as demonstrated in \citet{banzatti24}. 

As a constraint on the observed emitting radii, we can also use the measured broadening of MIRI lines as demonstrated in \citet{banzatti24}. In EX~Lup, we find that only the higher-energy lines around 6000~K show evidence of being broader than the MIRI resolving power, with FWHM $\approx 170$~km/s, while lines from energies $< 4500$~K are unresolved. The radial emitting region corresponds to $\approx$~0.02--0.1~au for the 6000~K water and CO transitions from the line list presented in \citet{banzatti24}, out to a lower limit of $\gtrsim 0.2$~au for the lower-energy water transitions. These regions match very well what previously measured in EX~Lup from the broad (0.02--0.1~au) and narrow ($\gtrsim 0.2$~au) CO components as observed at the much higher resolving power of 95,000 with CRIRES on the VLT in \citet{banz15}.

\subsection{Molecular variability in EX Lup as compared to other disks}\label{sec: mol_variab}

To analyze the variability in molecular emission from the disk of EX Lup, we measured the line luminosity of multiple species observed in the IRS and MIRI epochs and compare them to those previously measured from IRS spectra in a sample of 60 T~Tauri disks \citep{banz20}. For the organic molecules, we measure the line flux in EX~Lup spectra over the same spectral ranges used in \citet{banz20} as shown in Figure \ref{fig: all_spectra} (vertical dashed lines). These ranges minimize contamination from other molecules, with the exception of the SP2 epoch (2008 outburst) where water emission increased across the whole spectrum and the organics are not convincingly detected (see Section \ref{sec: slab_model}). To be consistent with the procedure used in \citet{banz20}, we subtract the water models shown in Figure \ref{fig: all_spectra} from each epoch before measuring the organic fluxes.  As for water, we measure the total flux of the strong water lines at 17.1--17.4\,$\mu$m for comparison to previous work with IRS. These lines are reported in Table \ref{tab: line fluxes}.

Figure \ref{fig: IRS_correl} shows line luminosities measured in the four EX~Lup epochs as a function of accretion luminosity, in comparison to the sample from \citet{banz20}. For \ce{H2O} and the organics, EX~Lup has line fluxes that are within the spread of data from the broader sample, with the exception of the organics in outburst whose upper limits are definitely sub-luminous. In relative terms, though, it is clear that while \ce{H2O} emission increased in EX~Lup during outburst, it increased more in the low-energy lines than in the high-energy ones (Section \ref{sec: cooler}). 

\begin{deluxetable}{l c c c c}
\tabletypesize{\small}
\tablewidth{0pt}
\tablecaption{\label{tab: line fluxes} Blended line fluxes used in Figure \ref{fig: IRS_correl}.}
\tablehead{\colhead{Species} & \colhead{SP1}  & \colhead{SP2} & \colhead{JW1} & \colhead{JW2}\\
 & \multicolumn{4}{c}{($10^{-14}$ erg s$^{-1}$ cm$^{-2}$)}}
\tablecolumns{5}
\startdata
HCN & 5.74 (0.29) & 0.66 (1.25) & 5.44 (0.01) & 4.60 (0.01) \\
\ce{C2H2} & 2.55 (0.21) & 0.07 (0.90) & 2.45 (0.01) & 2.22 (0.01) \\
\ce{CO2} & 1.32 (0.16) & 0.66 (0.72) & 1.13 (0.01) & 1.22 (0.01) \\
\ce{H2O} & 4.50 (0.50) & 9.96 (2.00) & 3.97 (0.01) & 3.02 (0.01) \\
\enddata
\tablecomments{Line fluxes are measured over the spectral ranges indicated in Section \ref{sec: mol_variab} and Figure \ref{fig: all_spectra} for comparison with previous fluxes measured from Spitzer spectra, including the 17~$\mu$m line cluster for \ce{H2O}.}
\end{deluxetable}

\section{Discussion} \label{sec: disc}
\subsection{Accretion variability as a source of luminosity scatter in molecular spectra}

The infrared line luminosity of multiple molecules has been found in previous work to correlate strongly with accretion luminosity \citep{salyk11_spitz,banz20,banz23,banzatti24}, suggesting a key role for UV radiation in inner disk heating and gas excitation as expected by models \citep[e.g.][]{walsh15,najita17,anderson21}. Still, these correlations have scatter of about one order of magnitude at any given accretion rate (e.g. Figure \ref{fig: IRS_correl}). Such large scatter is not uncommon in correlations found with accretion rates \citep[e.g.][]{manara22}, and it is likely amplified by multiple factors that affect accretion estimates, including the uncertainties in $A_{\rm{V}}$ (see Section \ref{sec: accr}) and different methods used in different works to estimate accretion (Balmer Jump or line luminosity relations). 

One question we wish to address in this work is to what extent accretion variability could contribute to the luminosity scatter observed in infrared molecular emission from disks \citep{cn11,salyk11_spitz,MINDS23} through variable disk heating and UV irradiation. A key problem to keep in mind is that, typically, the accretion luminosity has not been measured simultaneously to an infrared spectrum. In the context of correlations that use literature values for accretion as shown in Figure \ref{fig: IRS_correl}, this means that the value shown on the y axis (the measured molecular luminosity) does not correspond in time to the value on the x axis, which could have varied between the two datasets by factors of 2--10 following the typical accretion variability and short-term episodic bursts of T~Tauri stars \citep[][]{fischer23}. 

\begin{figure*}
\centering
\includegraphics[width=0.85\textwidth]{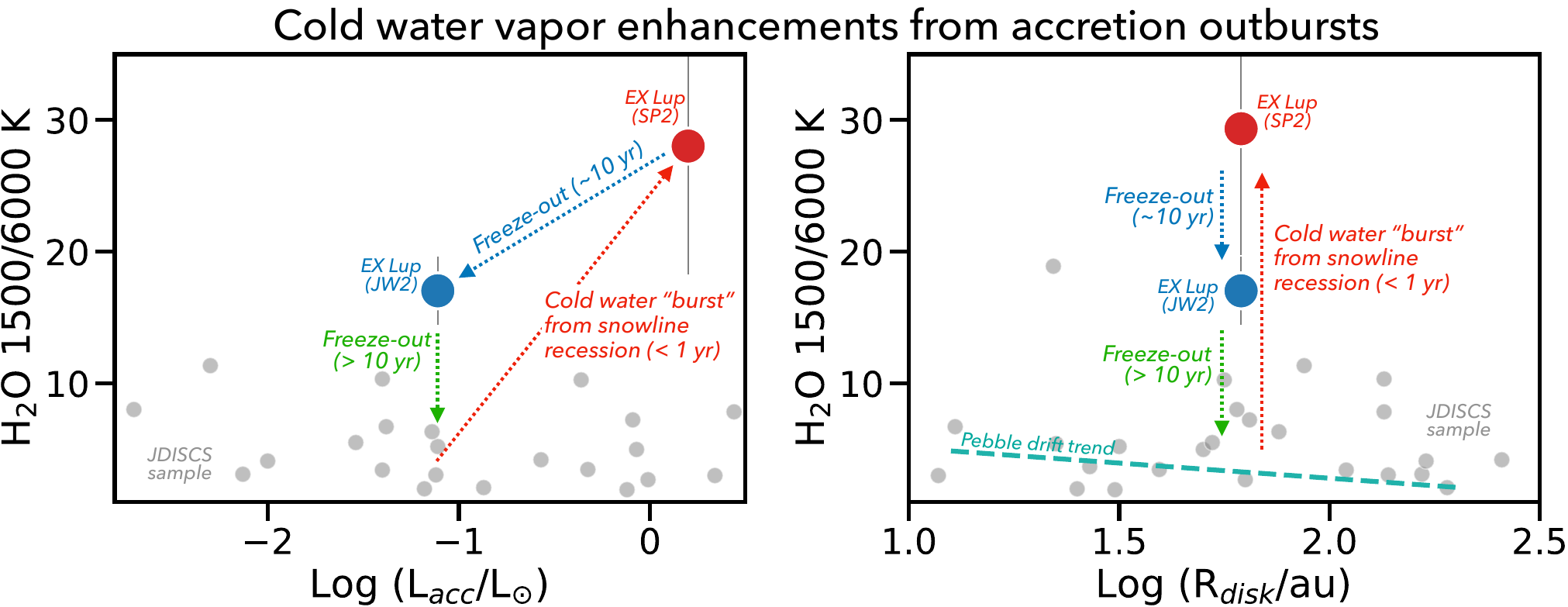} 
\caption{Cold water measured from the 1500/6000~K line ratio in the MIRI spectra of EX~Lup in comparison to the larger sample from the JDISC Survey \citep[grey datapoints from][]{banzatti24}. The dashed line in the right plot shows the anti-correlation between the line ratio diagnostic of cold water and the dust disk radius, attributed to pebble drift \citep{banz23b,banzatti24}. The large cold water excess measured in EX~Lup is suggestive of the role of accretion outbursts in triggering ice sublimation at the snowline (Figure \ref{fig: cartoon}), enriching the inner disk surface with $\sim 170$~K water vapor (Figure \ref{fig: 23umlines_ratios}). }
\label{fig: cool_excess}
\end{figure*}

The luminosity of hot molecular tracers ($> 400$~K) measured in EX~Lup is very similar across 20 yr (Figure \ref{fig: IRS_correl}) in the three quiescent epochs that had similar accretion rate, even after a strong accretion outburst that happened in between. On shorter timescales ($\lesssim$~month, since SP2 was taken about 2.5 months after the peak of the 2008 outburst), a strong accretion outburst can change molecular fluxes by a factor of a few and up to almost a factor 10, increasing or decreasing depending on the molecule (see Table \ref{tab: line fluxes} and Figures \ref{fig: IRS_correl}). While the sparse epoch sampling leaves large uncertainty on the exact timescales, this is consistent with general model expectations for short chemical equilibrium timescales in a UV-irradiated inner disk surface \citep[e.g.][]{thi05,najita17}. In EX~Lup, water emission varies along the global trend observed in the larger disk sample, while the organics decrease in outburst to the level of upper limits observed in other disks (Figure \ref{fig: IRS_correl}); the case of EX~Lup suggests that undetected organics in other disks might, in some cases, be due to phases of stronger UV photo-dissociation from an increased accretion luminosity. 

The molecular variability observed in EX~Lup demonstrates how much a factor of $> 10$ change in accretion luminosity (Table \ref{tab: epochs}) could contribute to the scatter observed at any given accretion rate in the larger sample of disks (Figure \ref{fig: IRS_correl}). Since the change in EX~Lup is at the higher end of the more typical accretion variability common to all T~Tauris \citep[2--10,][]{fischer23}, these results suggest that accretion variability can be a significant factor contributing to the observed scatters up to a factor of 2--5 in molecular luminosity. In other words, accretion variability is a hidden parameter that could produce part of the observed scatter in correlations between molecular spectra in large disk samples \citep[e.g.][and Figure \ref{fig: IRS_correl}]{banz20}, and it can produce a mismatch between the observed molecular spectra and accretion rates in case of non-simultaneous observations of these two datasets.

Besides considerations simply based on integrated molecular luminosities, the most important information to extract from the spectra would actually be how accretion variability changes the gas properties in terms of temperature and density as a function of disk radius. In this work, however, due to the much lower resolution of the IRS spectra and the large uncertainty in their analyses \citep[e.g.][]{salyk11_spitz,banz12} it was only possible to compare the MIRI-derived best-fit models to the IRS spectra, finding that the organic spectra are approximately the same during quiescent epochs across 20 yr. More distinct and clear is the change in water emission during and after the 2008 outburst, which we discuss in more detail in the following sections.

\subsection{A cold water vapor ``burst" from ice sublimation}
In Figure \ref{fig: cool_excess}, we compare the \ce{H2O} 1500/6000~K line flux ratio, a diagnostic of the cold water reservoir (Figure \ref{fig: 23umlines_ratios}), as measured in EX~Lup to values measured in the disk sample in \citet{banzatti24}.
In the left panel of Figure \ref{fig: cool_excess}, we visualize its evolution as a function of accretion phase in EX~Lup. In the SP2 outburst epoch this ratio is very high, and at the lower accretion rate of the MIRI epochs, 15 yr after the outburst, the ratio is still higher than in most other disks, indicative of an unusually large cold water reservoir (Section \ref{sec: cooler}).

In the right panel of Figure \ref{fig: cool_excess}, we illustrate the same line ratio as a function of dust disk sizes R$_{disk}$ measured from ALMA, in reference to the trend attributed to pebble drift where drift-dominated (i.e. smaller) disks have a relatively larger cold water reservoir \citep{banz20,banz23b,banzatti24}. 
While a more efficient pebble drift should increase the cold water vapor reservoir and in turn the measured 1500/6000~K ratio, the disk radius of EX~Lup \citep[from][]{Ansdell18} places it in the middle of the trend, thus offering no explanation for its unusually high 1500/6000~K value just based on pebble drift.
Multiple processes have been proposed to affect the cold water reservoir \citep[see discussions in][]{banzatti24,gasman25}, including: dynamical interactions in wide binary systems \citep{grant24}, dust depletion in inner disk cavities \citep{perotti23}, and the pebble drift dependence on time and on gap locations and depths \citep{kalyaan21,kalyaan23,easterwood24,mah24}. However, EX~Lup does not fit into any of these explanations: it is not part of a wide binary system, it does not have a large inner dust cavity\footnote{A potential small dust cavity of 0.2--0.3~au was proposed from the SED fit by \citet{juhasz12}, but CRIRES spectra show broad CO lines emitting from this region \citep{banz15}.}, and it does not show evidence for disk gaps down to a resolution of 0''.3 \citep{hales18}. 

\begin{figure}
\centering
\includegraphics[width=0.47\textwidth]{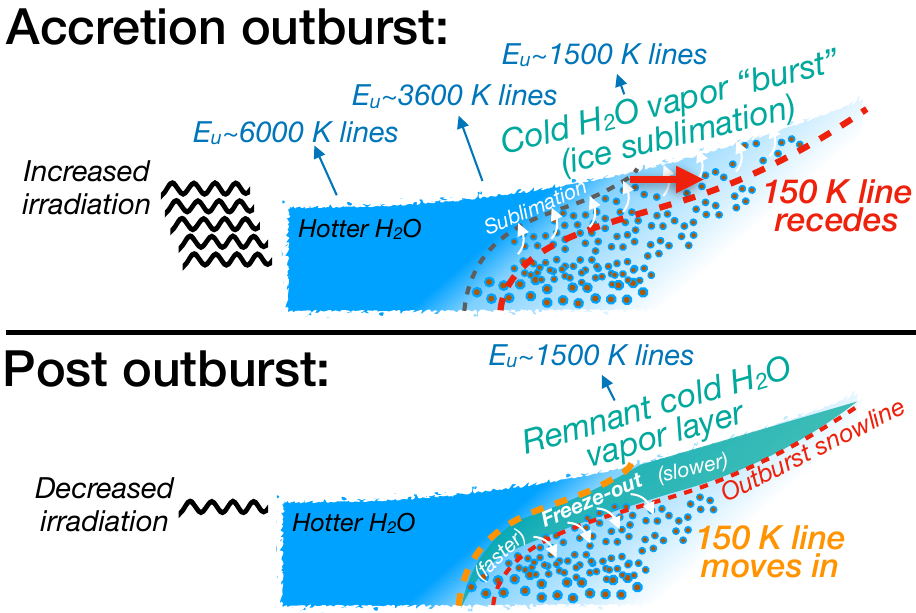} 
\caption{Illustration of the proposed interpretation for the water emission variability observed in EX~Lup during and after accretion outbursts. The increased irradiation and heating of the inner disk during outburst is proposed to push the snowline to larger radii, triggering a burst of cold water vapor from ice sublimation. A remnant layer of cold water vapor is left behind due to the long freeze-out timescales in the disk surface after outburst, to explain the increased cold water excess in the MIRI epochs analyzed in this work.}
\label{fig: cartoon}
\end{figure}

What, then, may be the origin of the unusually large cold water reservoir in EX~Lup? Its recent accretion history suggests one additional process that may increase the cold water diagnostic ratio shown in Figure \ref{fig: cool_excess}: episodic events of ice sublimation triggered by accretion outbursts (Figure \ref{fig: cartoon}).
The increase in water emission during the 2008 outburst suggested that a larger disk area became water-rich \citep{banz12}, following the increased heating of the inner disk \citep{abraham09,juhasz12}. A potential explanation that was proposed is a recession of the water snowline to larger radii during the outburst, as expected by models \citep[e.g.][]{vorobyov22}, causing the sublimation of a large ice reservoir previously located beyond the quiescent snowline. 
\rev{By modeling the disk spectral energy distribution (SED), we now estimate that the midplane snowline radius could have been pushed out by a factor of $\approx 2$ during the 2008 outburst in EX~Lup (Appendix \ref{app: model}).}

In this work, we confirm evidence for a water snowline recession from the larger increase of low-energy lines relative to high-energy lines in 2008 (Section \ref{sec: cooler}), which is consistent with increased emission at temperatures down to ice sublimation ($< 180$~K, Figures \ref{fig: all_spectra} and \ref{fig: 23umlines_ratios}). EX~Lup therefore provides the first example of a water ice sublimation ``burst" from a receding snowline caused by an accretion outburst in a T~Tauri disk (Figure \ref{fig: cartoon}). A water snowline recession has previously been inferred only in the more extreme case of protostellar and FUor outbursts, and not directly from water observations \citep{cieza16,hsieh19}. 

\begin{figure*}
\centering
\includegraphics[width=0.9\textwidth]{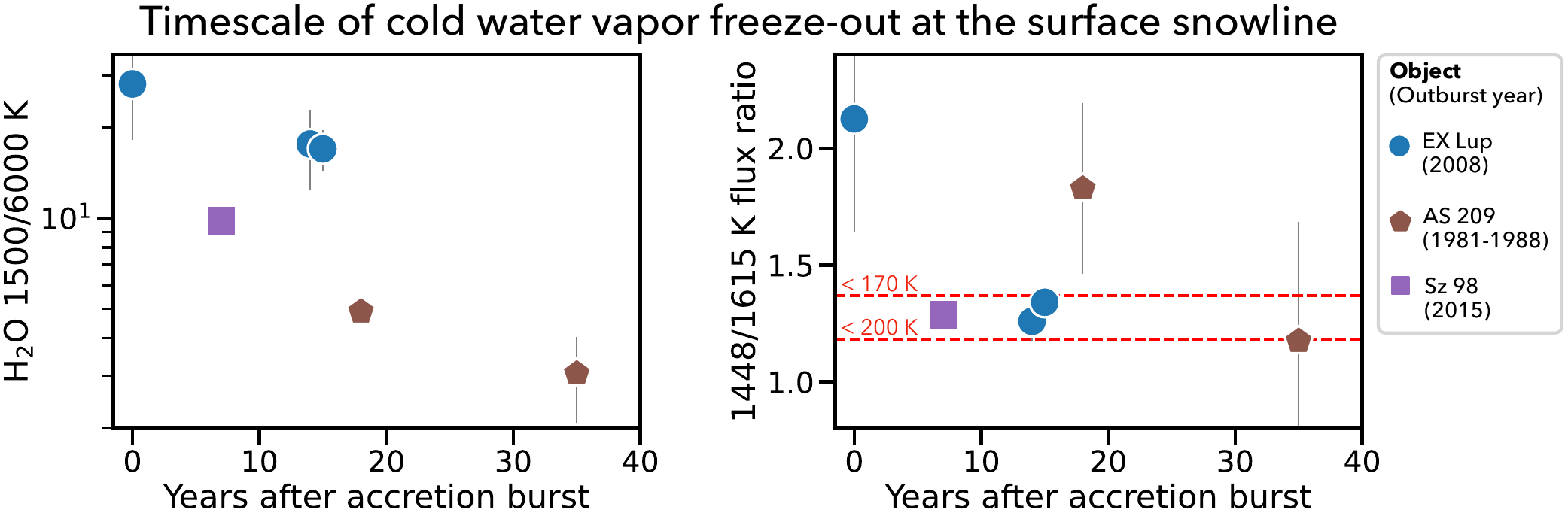} 
\caption{\rev{Time evolution measured in the cold water lines tracing the surface snowline. Both the cold water area relative to the inner hot water (diagnostic to the left) and the cold water temperature (diagnostic to the right) evolve significantly over 10--30 years after accretion outbursts. Two additional systems with evidence for recent accretion bursts are included in this figure: AS~209 and Sz~98 (see Section \ref{sec: time evolution} for details).}}
\label{fig: Burst_timescale}
\end{figure*}

\subsection{Cold water lines as probes of the accretion history and freeze-out timescales at the surface snowline} \label{sec: time evolution}
Once the inner disk cools down after outburst, the surface layer that was enriched in cold water vapor by ice sublimation during outburst is once again under freeze-out temperature and the ice reservoir can start to build up again (Figure \ref{fig: cartoon}). However, \rev{the freeze-out timescales are much longer than the sublimation timescales, and it should take 1--1000 years to turn the cold water vapor into ice again depending on the gas density, with longer times in protostellar envelopes and shorter in protoplanetary disks} \citep[e.g.][]{visser15,rab17,jorgensen20,houge23,ros24,lee25}. \rev{The MIRI spectra of EX~Lup may allow us for the first time to directly measure the water freeze-out timescale in Class II inner disk surfaces.}

In this regard, we should consider two potential origins for the ice-sublimation ``burst" in EX~Lup: the strong outburst in 2008 or the more moderate burst in March 2022 (Figure \ref{fig: light-curve}). Given the much smaller amplitude and duration of the 2022 burst \citep[$\sim3$ times shorter and $\sim10$ times weaker accretion than the 2008 outburst,][]{cruzsaenzdemiera23,wang23}, the most likely trigger of a large mass of sublimated ice was the stronger outburst in 2008, implying $> 10$~yr freeze-out timescales for the excess cold water in the disk surface. Given the similarity of the JW1 and JW2 spectra, we can also conclude that no significant depletion has happened over 7 months in between them, again supporting $> 1$~yr long freeze-out timescales. This timescale for re-adsorbing excess cold water after an accretion burst matches model expectations: \rev{using Equation 12 from \citet{rab17} and all the standard values assumed there for the dust and gas properties, the freeze-out timescale at 150~K in the disk surface (using $10^{9}$~cm$^{-3}$ for the total hydrogen number density $n_{<\rm{H}>}$) with a gas-to-dust ratio in the range 100-1000 (to account for some dust settling) is 2--20 yr.}
In a few more years, the cold water vapor in EX~Lup might further decrease to the values observed in other disks (downward arrow in Figure \ref{fig: cool_excess}).

With freeze-out timescales that are several years long, it should be possible to observe excess infrared cold water emission in inner disks that have had recent accretion bursts, similarly to what has been done with other gas tracers for protostellar outbursts \citep{hsieh19,anderl20}. In other words, while the hotter gas reservoir traced by the higher-energy lines should more closely reflect the current accretion luminosity that is irradiating the inner disk \citep{banz23b,banzatti24}, the colder gas closer to the ice sublimation front may have a delay with the current accretion state and still reflect recent phases of higher accretion. \rev{We explore this idea in Figure \ref{fig: Burst_timescale}, where we show the cold water vapor diagnostics as a function of time after the accretion outburst. We include EX~Lup as measured in this work in the SP2 and MIRI epochs, as well as two more disks that have published MIRI spectra and where we could find evidence for an accretion outburst ($\Delta L_{\rm{acc}} > 10$) in the past 30--40 years.}

One of these systems is AS~209, which has been found to show an unusually strong cold water spectrum that decreased in 17 years by a factor of $\sim3$ between a Spitzer spectrum in 2006 and a MIRI spectrum in 2023 \citep{munozromero24}. The absence of continuum variability in AS~209 between the two spectra led \citet{munozromero24} to exclude the accretion burst scenario; however, now EX~Lup demonstrates that cold water can survive in the disk surface for more than 10 yr after an outburst, suggesting the possibility that a strong accretion outburst might have happened in AS~209 up to several years before the IRS epoch was taken. In fact, the dust will cool down to pre-outburst levels much faster than the gas \citep[e.g.][]{vorobyov22}, where instead the long freeze-out timescales can leave signatures for several years (see above). \rev{By searching the literature, we find that the accretion luminosity and rate in AS~209 dropped by more than a factor of 10 from log$L_{\rm{acc}}/L_{\odot} \sim 0.5-1$ in 1981 and 1988 \citep[from UV excess measured in][]{valenti93,Johns-Krull00} down to log$L_{\rm{acc}}/L_{\odot} \sim -1.12$ in 2008 \citep[from optical lines measured in][]{fang18}, with the same value of extinction ($A_V = 1.1$~mag). We could not find additional accretion measurements in between these epochs, therefore the exact time and duration of the outburst are uncertain. In Figure \ref{fig: Burst_timescale}, we assume the burst ended in 1988, the last epoch that gives evidence for it.}

\rev{The second system we include is Sz~98, which has been found to have an anomalously strong cold water emission spectrum in \citet{gasman23,gasman25}. \citet{gasman23} reported the surprising finding that HI is not detected in the 2022 MIRI spectrum (giving an upper limit of log$L_{\rm{acc}}/L_{\odot} \sim -2$, \citep{Tofflemire25}), while a high accretion luminosity of log$L_{\rm{acc}}/L_{\odot} \sim -0.7$ was previously reported in 2015 data by \citet{alcala17,alcala19}. The accretion behavior of this system has been observed in other works using data from 2008--2009, all reporting a similar log$L_{\rm{acc}}/L_{\odot}$ in the range of -1.5 and -1.2 \citep{antoniucci14,fang18}. Even in this case, a similar extinction of $A_V = 1-1.25$~mag was used \citep{alcala17,fang18}, therefore the different accretion cannot be attributed to that. Based on its large photometric variability, \citet{antoniucci14} even suggested the potential ``EXOr" nature of Sz~98. The large change measured in accretion between 2008, 2015 (about a factor of 10 higher), and the recent 2022 MIRI spectrum (more than a factor of 10 lower), supports a scenario where this system had an accretion burst in 2015, which is the date we assume in Figure \ref{fig: Burst_timescale}. With only one epoch available to date, the infrared spectrum of Sz~98 should be re-observed in the future to test if its cold water reservoir is reducing with time.}

\rev{A potential way to identify enhanced cold water due to recent accretion outbursts may be the region defined by EX~Lup in the diagnostic diagrams shown in Figures \ref{fig: diagnostic_diagram} and \ref{fig: 23umlines_ratios}.} Based on that, an accretion burst origin could be considered for the unusually large cold water reservoir recently discovered in the young disk of IRAS~04385-2550 \citep[][]{banzatti24}, the only object in the whole JDISCS sample to match the cold water line ratio observed in EX~Lup (Figure \ref{fig: 23umlines_ratios}). Its accretion history is not recorded, but evidence for multiple ejection events (possibly driven by accretion outbursts) has been found in the HH~408 outflow associated to this star \citep{bally12}, which has recently shown stochastic variability with amplitude similar to bursting sources \citep{cody22}.

\begin{figure}
\centering
\includegraphics[width=0.45\textwidth]{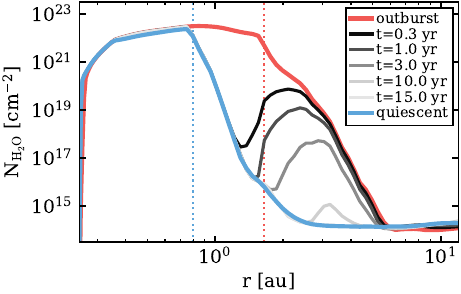} 
\caption{\rev{Post-outburst freeze-out of cold water vapor produced during an accretion outburst. The modeling framework is based on \citet{rab17} and described in Appendix \ref{app: model}. The red solid line shows the water column density at the end of the outburst, in reference to the quiescent phase in blue. The solid grayscale lines show the evolution at different times after the outburst, indicating $> 3$~yr freeze-out timescales for cold water after an outburst. The vertical dotted lines show the midplane water snowline in quiescence (blue) and outburst (red).}}
\label{fig: Freezeout_model}
\end{figure}

\rev{Lastly, in Figure~\ref{fig: Freezeout_model}, we show a prediction for the post-outburst evolution of the cold water vapor from a 2D radiation thermo-chemical model adapted to EX~Lup (described in Appendix \ref{app: model}). The models show the recession of the water snowline during outburst, increasing the water column density in the disk surface at 1--5~au. The midplane snowline (vertical dotted lines) quickly resets to its quiescent location after outburst; however, in the disk surface at a few au the freeze-out is much slower due to the lower densities, resulting in a cold water vapor reservoir that survives for $> 3$~yr after the outburst. This cold, radially extended reservoir matches the emitting region of the cold water vapor resolved with MIRI in other disks and proposed to trace the surface snowline \citep{banzatti24}, consistent with the region shown by the thermo-chemical model (Appendix \ref{app: model}). The MIRI observations of EX~Lup show uncommon cold water after $> 10$~yr, which seem to imply sligthly longer freeze-out timescales than the model predicts. One factor ignored in the model is the moderate burst that happened in 2022, which could have delayed the water freeze-out in EX~Lup. A promising direction for future work is to perform a more comprehensive thermo-chemical modeling of the evolution of EX~Lup and other outbursting Class II objects, including line emission spectra from all molecules.}

\subsection{Implications for planet formation} 
A recession of the water snowline should have significant implications on planet formation. Since water ice increases the growth of solids from grains to planetesimals, the snowline is recognized as a fundamental location for planet formation \citep{ros13,drazkowska17}. A sudden recession of the snowline would expose a large mass of icy solids to sublimation and alter dust properties and growth, which could slow down or even halt planet formation \citep{houge23,houge24}. Multiple bursts causing continued recessions of the snowline could even permanently hinder the growth to large planet cores, leaving only small and dry rocky planets to form. On the other hand, freeze-out, especially if slow, could facilitate the growth of large pebbles that can trigger the streaming instability and form ice-rich planetesimals \citep{ros24}. Moreover, the release of a large water vapor mass within the snowline would decrease the C/O ratio in the inner disk gas, and therefore cause an oxygen-rich chemistry in the atmosphere of gas-accreting planet cores that may have already formed. 

Monitoring ice sublimation during accretion outbursts in Class II disks would have great significance for theories of planet formation, as it would enable the calibration of models on the ice reservoir at the snowline in terms of its mass, grain properties, and the role of accretion variability in affecting core formation and the following planet atmospheric chemistry.

\section{Summary \& Conclusions} \label{sec: concl}
In this work, we have analyzed two epochs of JWST-MIRI spectra of molecular emission from the disk of EX Lup observed in 2022 and 2023, about 15 years after a strong accretion outburst in 2008. We have compared the observed molecular emission to two previous epochs from Spitzer-IRS in 2005 (quiescence) and 2008 (during the outburst), as well as to a large sample of Class II disks observed previously with IRS or MIRI.
The main findings and conclusions of this work are:

\begin{itemize}
    \item the mid-infrared molecular emission observed in EX Lup with MIRI in 2022 and 2023 is consistent with quiescent Spitzer spectra from 2005, with a factor of $<2$ change in molecular line fluxes; the organic molecules that disappeared during the 2008 outburst are now back and consistent to pre-outburst levels;

    \item a re-analysis of water emission based on recent diagnostics established from MIRI spectra shows that EX~Lup had a cold water vapor ``burst" in 2008; we interpret this as confirming evidence for a recession of the snowline causing the sublimation of part of the ice reservoir from beyond the quiescent snowline. This is the first time such a process is captured in a Class II disk; similar effects have been observed before in protostellar envelopes and the stronger FUor outbursts only;

    \item in comparison to other disks observed with MIRI, EX~Lup now shows an unusually strong cold water emission; we propose that this is the remnant of the cold water vapor ``burst'' from ice sublimation triggered during outburst; this implies that freeze-out timescales in inner disk surfaces are several years long, matching modeling expectations; the infrared cold water lines may retain the imprint of accretion bursts in Class II disks up to 10--20~yr, providing a way to study their chemical evolution as a function of accretion history;

    \item the molecular variability observed in EX Lup across different accretion epochs covers a significant portion of the spread along accretion trends observed in disk samples before; accretion variability can play a significant role in changing the observed organic-to-water line ratios, in addition to increasing the cold water reservoir;

    \item while this work adds evidence for accretion variability to play a significant role in inner disk chemistry and possibly planet formation, the results were limited by the different resolving power of IRS and MIRI-MRS and a sparse epoch sampling; this highlights the need for future monitoring of Class II objects to determine the detailed heating, cooling, chemical, sublimation, and freeze-out timescales in inner disks.
\end{itemize}

Future modeling work should investigate the scenario proposed in this work and its implications for planet formation, in particular how much water ice can be sublimated during an outburst as a function of its strength and duration. While models have explored outer disks and envelopes around protostars with stronger outbursts \citep[e.g.][and references therein]{molyarova18,jorgensen20}, the regime of EXor outbursts and other accretion bursts in the Class II phase still need a systematic investigation of their inner disk molecular variability, which EX~Lup and future dedicated programs could support.

\acknowledgments
The authors acknowledge helpful discussions with S. Krijt, A. Houge, and T. Bergin on sublimation and freeze-out timescales in inner disks, as well as feedback from an anonymous referee that helped improve this work significantly.
This work is based on observations made with the NASA/ ESA/CSA James Webb Space Telescope. The JWST data used in this paper can be found in MAST:
\dataset[10.17909/rjt1-jd72]{http://dx.doi.org/10.17909/rjt1-jd72}.
The data were obtained from the Mikulski Archive for Space Telescopes at the Space Telescope Science Institute, which is operated by the Association of Universities for Research in Astronomy, Inc., under NASA contract NAS 5-03127 for JWST. The observations are associated with JWST GO Cycle 1 programs 2209 and 4427.
This work included observations collected at the European Organization for Astronomical Research in the Southern Hemisphere under ESO program 110.24BN.001

The authors acknowledge support from NASA/Space Telescope Science Institute grants JWST-GO-2209 and JWST-GO-4427. This work was also supported by the Hungarian NKFIH projects K-132406, K-147380, and ADVANCED 149943.
This work was partly supported by the NKFIH excellence grant TKP2021-NKTA-64.
FCSM received financial support from the European Research Council (ERC) under the European Union’s Horizon 2020 research and innovation programme (ERC Starting Grant “Chemtrip”, grant agreement No 949278).
CFM was partly funded by the European Union (ERC, WANDA, 101039452). Views and opinions expressed are however those of the author(s) only and do not necessarily reflect those of the European Union or the European Research Council Executive Agency. Neither the European Union nor the granting authority can be held responsible for them.
CHR acknowledges the support of the Deutsche Forschungsgemeinschaft (DFG, German Research Foundation) Research Unit ``Transition discs'' - 325594231. CHR is grateful for support from the Max Planck Society.

\software{
Matplotlib \citep{matplotlib}, NumPy \citep{numpy}, SciPy \citep{scipy}, Seaborn \citep{seaborn}, Astropy \citep{astropy:2013, astropy:2018, astropy:2022}, LMFIT \citep{lmfit}, iSLAT \citep{iSLAT,iSLAT_code}, iris \citep{iris}.
}

\newpage

\appendix

\section{Updated MIRI spectra reduction} \label{app: reduction}
All MIRI data reduction of was performed with the JWST pipeline \citep[see][]{Bushouse2024} version 1.13.4, and was based on the pmap1200 calibration scheme. 
We followed for most of the pipeline stages the recommended steps and parameter settings. Only for the spectral cube building and spectral extraction we used slightly different settings. All the observations were performed using a 4-point dither pattern optimized for point sources. Due to the pointing offset in the first JWST epoch, two of the dither positions where outside of the Field Of View (FOV) of the short wavelength module of MRS, leading the a total loss of half the data at the shorter wavelengths. Compared to the original publication of the first data set, the improved geometrical distortion correction \citep{Patapis2024} using the construction of the spectral cubes \citep{Law2023} substantially improved the data quality of the spectral cubes constructed from the two highly offset usable dither positions.  We relied on the extract1D step to extract the 1D spectrum from the 3D spectral cubes. We used an extraction aperture of 1.5 times the Full Width Half Max (FWHM) of the Point Spread Function (PSF), which gives slightly higher signal-to-noise than other extraction apertures for point sources. For both observations, subtraction of the infrared background was done at the  extract1D step, where the background was estimated by a annulus around the source position outside of the extraction aperture. 
Spectral de-fringing was applied to all data sets, both at the Spectroscopy~2 pipeline stage as well on the extracted 1d spectra (see Crrouzet et al. 2025 (submitted), for further details on the de-fringing). Compared to the \cite{kospal23} publication, an improved time dependent flux calibration is applied \citep{gasman2023, Law2025}, which has an estimated photometric  uncertainty between repeated observations of better than 1~\% in the 5 to 18~$\mu$m wavelength range.  The wavelength calibration of the final spectra is estimated to be accurate within 9 km s$^{-1}$ at 5 $\mu$m and 27 km s$^{-1}$ at 28 $\mu$m \citep{Labiano2021, Argyriou2023}.

\section{The uncommon cold water spectrum from the inner disk of EX~Lup} \label{app: MIRI_compar}

\begin{figure*}
\centering
\includegraphics[width=1\textwidth]{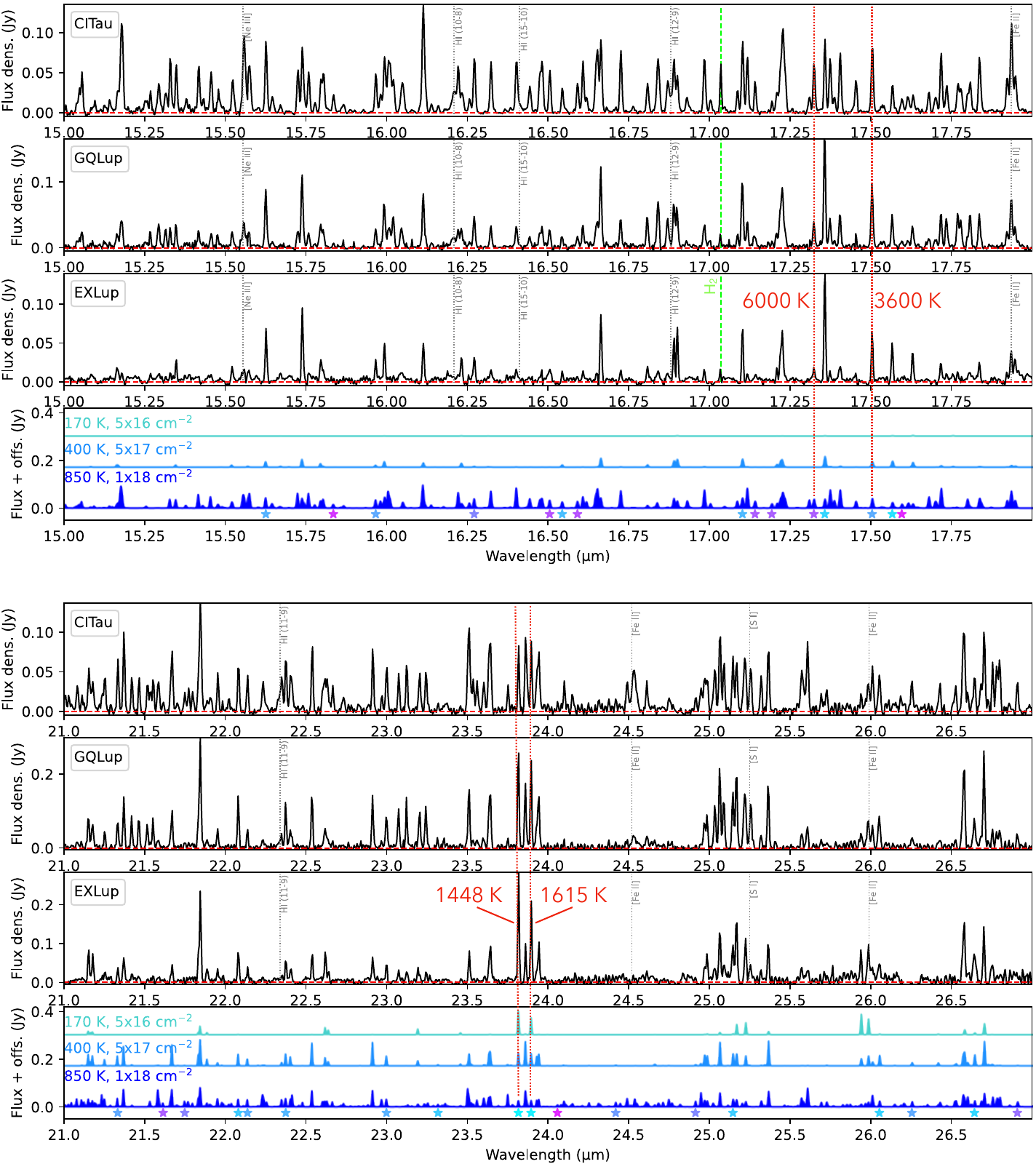} 
\caption{Water emission in EX~Lup as compared to that observed with MIRI in other representative disks, following Figure 8 in \cite{banzatti24}. A portion of the short wavelengths (top) and long wavelengths (bottom) illustrates that the spectrum in EX~Lup is dominated by lower-energy lines that get excited at intermediate to low temperatures down to ice sublimation, while the spectra of GQ~Lup and CI~Tau include increasing flux from hotter emission. Individual slab models and the un-blended transitions (marked with stars) presented in \cite{banzatti24} are shown at the bottom in each plot. \rev{The diagnostic lines introduced in \cite{banzatti24} and used in Figures \ref{fig: diagnostic_diagram} and \ref{fig: 23umlines_ratios} in this paper are marked in red and included in Table \ref{tab: diagnostic fluxes}.}
}
\label{fig: MIRI_spec_compar}
\end{figure*}

Figure \ref{fig: MIRI_spec_compar} shows water emission in EX~Lup as compared to that observed with MIRI in other disks following \cite{banzatti24}. 
In comparison to the two disks used as representative examples in \citet{banzatti24}, CI~Tau (dominated by a hot, $\sim 850$~K component) and GQ~Lup (which instead includes strong emission from colder water at larger disk radii), EX~Lup is dominated by lower-energy lines that can be recognized by comparison to the slab models shown at the bottom of the figure.
\rev{The specific transitions used as diagnostics of the water reservoir at different temperatures in Figures \ref{fig: diagnostic_diagram} and \ref{fig: 23umlines_ratios} in this paper, approximating a temperature gradient from the inner disk out to the water snowline \citep{munozromero24b,banzatti24}, are marked and labeled in Figures \ref{fig: MIRI_spec_compar} and \ref{fig: EXLup_deblended} and reported in Table \ref{tab: diagnostic fluxes}. The 6000~K line near 17.32~$\mu$m is most sensitive to the hotter region represented by the 850~K model, the 3600~K line near 17.50~$\mu$m is still prominent in the warm region represented by the 400~K model, while the two $\sim 1500$~K lines near 23.85~$\mu$m are most sensitive to the cold region near the ice sublimation front. The use and interpretation of these lines as diagnostics of the radial distribution of water in inner disks is demonstrated in detail in \citet{banzatti24}.
As demonstrated in Figure 11 in \citet{banzatti24}, the flux asymmetry in these two lines is very sensitive to the coldest temperature detected with MIRI, with the 1448~K line flux that increases relative to the 1615~K line flux at colder temperatures (visible by comparison to the slab models at the bottom of Figure \ref{fig: MIRI_spec_compar}).}
The overall spectral line energy distribution of the three spectra in this figure clearly illustrates the cooler emission in EX~Lup from the relatively more prominent lines at longer wavelengths, as discussed in \citet{banzatti24}.

\section{Water diagnostic lines measured from EX~Lup spectra} \label{app: deblending}

\begin{figure*}
\centering
\includegraphics[width=1\textwidth]{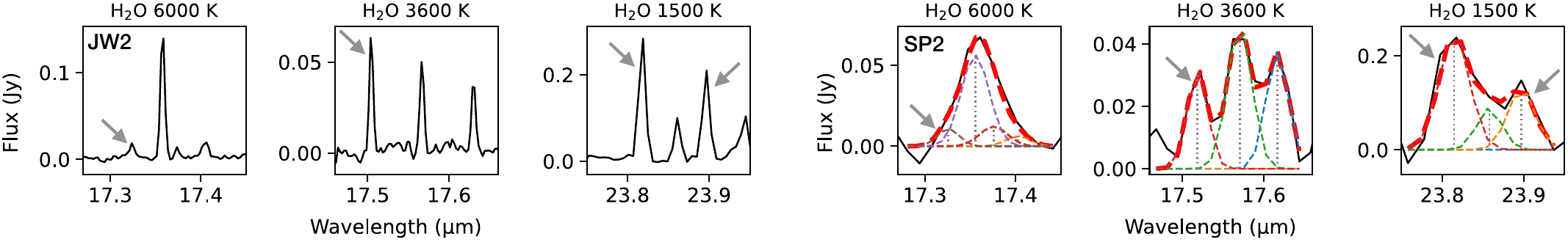} 
\caption{\rev{Diagnostic water lines defined in \citet{banzatti24} as observed in the spectra of EX~Lup. Left plots: JW2 spectrum observed with MIRI at the resolving power of R~$\sim 3000$, which resolves these lines from nearby transitions. Right plots: the same lines as observed in the SP2 spectrum with IRS (R~$\sim 700$); line de-blending is performed with iSLAT and shown with colored dashed lines, with a thicker red line showing the total fit (see Appendix \ref{app: deblending} for details).}
}
\label{fig: EXLup_deblended}
\end{figure*}

\rev{The diagnostic lines used in Figures \ref{fig: diagnostic_diagram} and \ref{fig: 23umlines_ratios} have been introduced in \citet{banzatti24} for water spectra as observed with MIRI, which has enough resolving power to resolve them from other nearby lines. These lines as observed in the MIRI spectrum of EX~Lup are included in Figure \ref{fig: EXLup_deblended} and their line flux in Table \ref{tab: diagnostic fluxes}. In this work, we measure these lines also in the Spitzer-IRS spectrum during outburst, where they need to be de-blended from other lines. Figure \ref{fig: EXLup_deblended} shows the line de-blending results for the SP2 spectrum, using the ``Line De-blender" function in iSLAT. This function fits blended lines to extract the flux of individual transitions following methods originally developed in \citet{banz13}. In short, blended lines are fitted with a combination of gaussian functions where centroid and FWHM are set by the known line rest wavelength \citep[from HITRAN,][]{hitran20} and resolving power (set by the specific instrument in use) while the gaussian amplitude is the principal free parameter. Effectively, even line centroid and FWHM are left free to vary within a limited range to account for some uncertainty in wavelength calibration (0.002~$\mu$m) and resolving power (10\%) of the IRS. The fit is performed with LMFIT \citep{lmfit} and the results for each line are shown in Figure \ref{fig: EXLup_deblended} where the specific diagnostic transitions are marked with an arrow and compared to their higher-resolution observation with MIRI. The extracted line fluxes are more uncertain than in the higher-resolution MIRI spectra but still provide estimates of the diagnostic line ratios used in this work and are included in Table \ref{tab: diagnostic fluxes}. The 3340~K lines used as proxy for the column density in Figure \ref{fig: diagnostic_diagram} cannot be extracted from Spitzer spectra due to excessive line blending of transitions from water, HCN, and \ce{C2H2}.}

\begin{deluxetable}{l l c c c c c c}
\tabletypesize{\small}
\tablewidth{0pt}
\tablecaption{\label{tab: diagnostic fluxes} Water diagnostic lines measured in EX~Lup.}
\tablehead{\colhead{Line ID} & \colhead{$E_u$} & \colhead{Component} & \colhead{Wavelength} & \colhead{SP1}  & \colhead{SP2} & \colhead{JW1} & \colhead{JW2}\\
 & & & \colhead{($\mu$m)} & \multicolumn{4}{c}{($10^{-14}$ erg s$^{-1}$ cm$^{-2}$)}}
\tablecolumns{8}
\startdata
6000~K & 6052~K & hot & 17.32395 & 0.29 (0.22) & 0.32 (0.14) & 0.16 (0.02) & 0.16 (0.02) \\
3600~K & 3646~K & warm & 17.50436 & 0.71 (0.45) & 0.97 (0.30) & 0.47 (0.04) & 0.38 (0.01) \\
1500~K & 1448~K & cold & 23.81676 & 2.19 (0.24) & 6.10 (0.79) & 1.55 (0.10) & 1.59 (0.03) \\
1500~K & 1615~K & cold & 23.89518 & 0.95 (0.17) & 2.86 (0.54) & 1.23 (0.02) & 1.19 (0.03) \\
\enddata
\tablecomments{The diagnostic lines included in this table follow definitions introduced in \citet{banzatti24}. The line ID identifies the approximate upper level energy of each transition; the ``1500~K" line flux used in Figures \ref{fig: diagnostic_diagram} and \ref{fig: 23umlines_ratios} is the sum of the two transitions near 23.85~$\mu$m indicated with arrows in Figure \ref{fig: EXLup_deblended}, as in \citet{banzatti24}.}
\end{deluxetable}

\section{Modeling the sublimation and freeze-out evolution of water} \label{app: model}
\rev{In this work, we present a radiation thermo-chemical model for EX~Lup focused on the evolution of water vapor during and after the 2008 outburst. This model should be considered representative for now as it has not been optimized to fit all the existing data including line emission spectra. We adopt model parameters (disk structure, stellar spectrum, etc.) from previous works that have performed dust radiative transfer models to fit photometry and Spitzer spectra in the quiescent and outburst phases \citep{Sipos2009,juhasz12,Sicilia-Aguilar2015}. The main goal of this exercise is to test the evolution timescales of water vapor, and in particular, the freeze-out timescales in the disk surface, assuming parameters appropriate for the 2008 outburst of EX~Lup.} 

\rev{We use the 2D radiation thermo-chemical code ProDiMo \citep[PROtoplanetary DIsk MOdel\footnote{\url{https://prodimo.iwf.oeaw.ac.at/} Version v3.0.0},][]{woitke09,kamp10,Thi2011}, which solves for the dust temperature and disk radiation field, the chemistry, and the heating-cooling balance. ProDiMo is a well-established tool to model protoplanetary disk chemistry that has been tested and validated over a wide range of data from the UV to the radio (both spectroscopic and photometric); more details on the disk structure and chemical network can be found in \citet{woitke16} and \citet{Kamp2017}.}

\begin{figure}
\centering
\includegraphics[width=\textwidth]{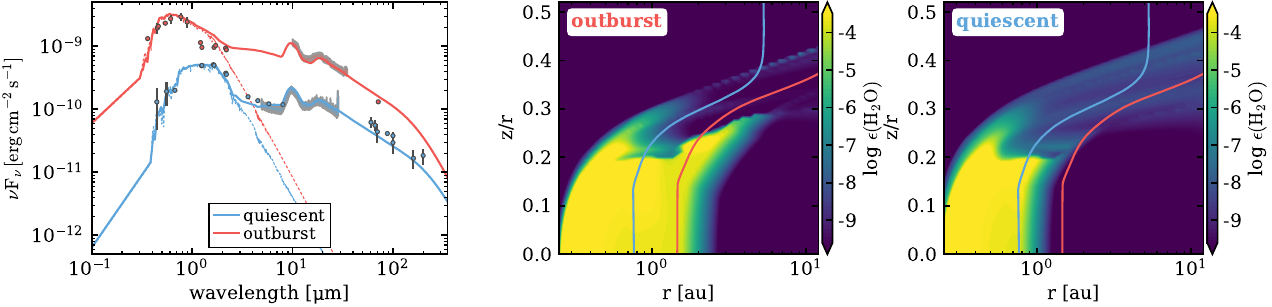} 
\caption{\rev{\textit{Left panel:} Stellar spectrum (dashed colored lines) and predicted SED (solid colored lines) during the quiescent (in blue) and outburst phase (in red) for the representative EX Lup model. The colored symbols with black borders show photometry measurements from the literature \citep{juhasz12}. The observed Spitzer spectrum (outburst) and JWST spectrum (quiescent), as used in this work, are shown in gray with stretched error bars for better visibility. \textit{Middle and right panels:} Water gas phase abundance (relative to the total hydrogen number density) at the end of the outburst and in the quiescent phase. The blue and red contours show $T_\mathrm{dust}=150\,\mathrm{K}$ in the quiescent and outburst phase, respectively. The water abundance between these two contours slowly decreases after the end of the burst until it reaches the initial (quiescent) level as shown in Figure \ref{fig: Freezeout_model}.}} 
\label{fig: model}
\end{figure}

\rev{To model the outburst in EX~Lup, we follow the approach presented in \citet{rab17} and similarly in \citet[][]{molyarova18}, where it is assumed that the outburst happens as an event of highly enhanced accretion rate onto the star, irradiating the inner disk with high-energy radiation. We first calculate the dust temperature and disk radiation field for the quiescent phase and evolve the chemistry for 1 Myr. The outburst is then modeled by adapting the stellar spectrum and the stellar luminosity, recalculating the disk dust temperature and radiation field, and then evolving the chemistry for half a year (approximately the outburst duration in 2008). After the end of the outburst, we reset the stellar properties to the quiescent state (see also Figure~\ref{fig: model}), recalculate the dust temperature and disk radiation field, and evolve the chemistry for another 300 years. For simplicity and to limit computational time, we assume that the gas temperature equals the dust temperature (i.e. we do not solve for the heating/cooling balance). For the scope of this work, this is a reasonable assumption as we focus on water chemistry, which is dominated by photo-processes and the freeze-out and desorption (photo-desorption, and thermal desorption) of water ice.}

\rev{In Figure \ref{fig: model} (left panel), we show the modeled SEDs for the quiescent and outburst phases. The modeled SEDs are consistent with the available photometry at 0.3--1000~$\mu$m \citep[adopted from][]{juhasz12} and match the Spitzer and JWST spectra available from the quiescent and outburst phases. We note that we do not change the disk structure during the outburst: the additional flux in the near/mid-infrared is a consequence of the increased stellar and accretion luminosity during the outburst. This additional optical and UV radiation is absorbed by the disk material, resulting in a hotter disk, which re-emits the radiation at longer wavelengths.}

\rev{The water abundance structure during quiescent and outburst phases is shown by the middle and right panels in Figure \ref{fig: model}. The enhanced cold water reservoir produced during the outburst at 1--5~au in the disk requires a few to 10 years after the end of the outburst to freeze out on dust grains. In the model presented here, the freeze-out timescale may be a factor of 3--5 faster in comparison to the observations in Figure \ref{fig: Burst_timescale}. The freeze-out timescale depends on various model parameters (see Sect. \ref{sec: time evolution}), but two factors that are more uncertain are the dust density, which may be too high in this region, and the water reservoir, which may be too deep in the disk; in both cases the higher density speeds up the freeze-out in the model. Another uncertain factor is the sticking coefficient, which may be different from the commonly assumed value of 1.}

\rev{Overall, this model is in qualitative agreement with the proposed scenario for the chemical evolution of EX Lup. More detailed modeling that includes direct comparison to the spectral lines will be useful to fully understand the chemical changes caused by outbursts and their consequences for the long-term chemical evolution of inner disks.}

\section{On organic chemistry in inner disks}
The chemistry of molecules in inner disk surfaces, including simple organics like \ce{HCN} and \ce{C2H2}, is regulated by the balance between formation and destruction reactions. The latter are dominated by UV photodissociation, causing lower organics abundances in disks with higher UV irradiation \citep{walsh15}. The emitting regions of \ce{HCN} and \ce{C2H2} are proposed to be closer-in and deeper towards the disk midplane in comparison to a more extended water surface layer \citep[e.g.][]{woitke24}. Our slab fit results are consistent with this scenario, by showing hotter and more compact emission from the organics relative to water. If the organics are photodissociated during outburst, it could be due to a deeper penetration of UV radiation into the disk, reaching areas of hydrocarbon emission. This has been previously proposed as an explanation for the disappearance of organics in the inner disk of EX Lup during the 2008 outburst \citep{banz12}. Another potential explanation is that, instead, organic emission is just hidden by stronger, optically thick emission from an inner hot water layer that has been observed to appear during outburst and disappear afterwards, leaving only a moderately-veiled stellar photosphere at 3~$\mu$m \citep{banz15}. This second scenario, however, still needs to be tested and validated by models.

Previous works have proposed or even detected the sublimation of ice-phase organics caused by increased disk heating during outburst \citep{molyarova18,lee19}, but they have focused on outer disk regions and more complex organics that are observed at millimeter wavelengths. In the hot inner disk organics in EX Lup we see the opposite effect: a decrease of organic emission during outburst. The inner disk organics, therefore, show an evolution that seems to be more connected to UV photo-chemistry and gas-phase formation, than to the sublimation of ices in the outer disk. Given the current knowledge and data, it seems reasonable to think that the inner disk organics have been dissociated by enhanced UV during the outburst and re-formed afterwards. However, as said above, all we can say is that in 2022 the organic emission is back and consistent with pre-2008-outburst levels. 
A future monitoring with higher-resolution observations should enable the distinction of these different scenarios from the change in the excitation, column density, and emitting area of the organic molecules.

The expected chemical timescales reacting to disk heating and UV irradiation in the warm molecular-emitting surface layers within a few au are as short as hours or days \citep{najita17,woitke16,semenov11}, but the strong dependence of timescales on gas density and temperature, which have large gradients in the molecular layers observed with MIRI, still leave large uncertainties on what should be expected to happen during and after UV bursts in disks of T~Tauri stars. Future modeling work should investigate inner disk molecules and attempt to explain their destruction and re-formation on timescales consistent with what was observed in EX~Lup.

\bibliography{EXLup_MIRI}{}
\bibliographystyle{aasjournal}

\end{document}